\documentclass[sigconf]{acmart}
\usepackage{subfigure}
\AtBeginDocument{%
  }

\copyrightyear{2026}
\acmYear{2026}
\setcopyright{cc}
\setcctype{by}
\acmConference[CHI '26]{Proceedings of the 2026 CHI Conference on Human Factors in Computing Systems}{April 13--17, 2026}{Barcelona, Spain}
\acmBooktitle{Proceedings of the 2026 CHI Conference on Human Factors in Computing Systems (CHI '26), April 13--17, 2026, Barcelona, Spain}
\acmDOI{10.1145/3772318.3790836}
\acmISBN{979-8-4007-2278-3/2026/04}

\newcommand{\dataset}{\textsc{ORG}}
\newcommand{\datasetfullname}{\textsc{Out-of-Reach Grasping}}

\begin{document}

\title{Point \& Grasp: Flexible Selection of Out-of-Reach Objects Through Probabilistic Cue Integration}

\author{Xuejing Luo}
\affiliation{
  \department{Department of Information and Communication Engineering, School of Electrical Engineering }
  \institution{Aalto University}
  \city{Helsinki}
  \country{Finland}}
  \affiliation{
  \institution{ELLIS Institute}
  \city{Helsinki}
  \country{Finland}}
\email{xuejing.luo@aalto.fi}

\author{Hee-Seung Moon}
\authornote{Corresponding author.}
\affiliation{%
  \department{School of Computer Science and Engineering}
  \institution{Chung-Ang University}
  \city{Seoul}
  \country{Republic of Korea}}
\email{hsmoon@cau.ac.kr}

\author{Christian Holz}
\affiliation{%
  \institution{ETH Z\"{u}rich}
  \city{Zurich}
  \country{Switzerland}}
\email{christian.holz@ethz.ch}

\author{Antti Oulasvirta}
\affiliation{
  \department{Department of Information and Communication Engineering, School of Electrical Engineering }
  \institution{Aalto University}
  \city{Helsinki}
  \country{Finland}}
  \affiliation{
  \institution{ELLIS Institute}
  \city{Helsinki}
  \country{Finland}}
\email{antti.oulasvirta@aalto.fi}

\renewcommand{\shortauthors}{Luo et al.}

\begin{abstract}
Selecting out-of-reach objects is a fundamental task in mixed reality (MR).
Existing methods rely on a single cue or deterministically fuse multiple cues, leading to performance degradation when the dominant cue becomes unreliable.
In this work, we introduce a probabilistic cue integration framework that enables flexible combination of multiple user-generated cues for intent inference.
Inspired by natural grasping behavior, we instantiate the framework with pointing direction and grasp gestures as a new interaction technique, \textsc{Point\&Grasp}.
To this end, we collect the \datasetfullname~(\dataset) dataset to train a robust likelihood model of the gestural cue, which captures grasping patterns not present in existing in-reach datasets.
User studies demonstrate that our selection method with cue integration not only improves accuracy and speed over single-cue baselines, but also remains practically effective compared to state-of-the-art methods across various sources of ambiguity. The dataset and code are available at \url{https://github.com/drlxj/point-and-grasp}.
\end{abstract}


\begin{teaserfigure}
  \centering
  \includegraphics[width=0.92\textwidth]{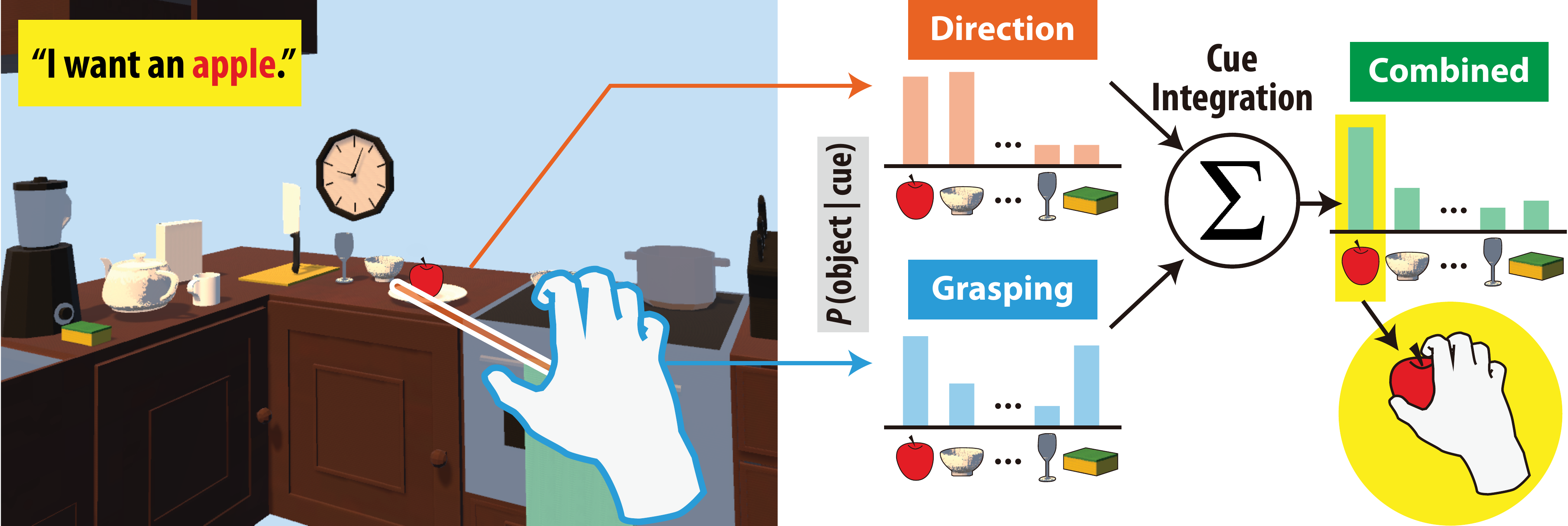}
  \caption{\textsc{Point\&Grasp} is a technique for selecting out-of-reach objects in virtual reality. While pointing direction suffers in cluttered scenes and grasping gestures become ambiguous when objects share similar shapes, \textsc{Point\&Grasp} combines both cues to enable more precise and usable target selection. 
  }
  \Description{The figure shows an example of a user in Virtual Reality selecting an object on a table. Several objects are present, including an apple, a cup, a plate, and a dish sponge. A virtual hand is extended in front of the table, forming a grasping gesture but not touching any object. Two types of cues are highlighted: directional pointing (shown with an arrow extending from the hand toward objects) and grasping gesture likelihood (based on how well the hand shape matches each object). Each cue produces a probability distribution over possible objects. On the right side, both cues are combined through an integration process, resulting in a single probability distribution where the apple has the highest likelihood. The combined inference determines that the apple is the correct target, illustrated with a hand selecting it.}
  \vspace{0.2cm}
  \label{fig:teaser}
\end{teaserfigure}


\maketitle

\section{Introduction} \label{sec: intro}

In mixed reality (MR) environments, users often need to interact with objects that lie beyond their physical reach. For example, a designer may want to grab a virtual tool floating above a 3D workspace, or a player may need to select a distant item in a game scene. Such 3D out-of-reach object selection is a fundamental task in MR interactions~\cite{bowman1997evaluation, argelaguet2013survey}. 
To support efficient and fluid selection, systems must accurately \emph{infer} user intent.
This is typically achieved by systems capturing behavioral signals produced as users act on their intent---in other words, user-generated \emph{cues}.\footnote{In the broader HCI literature, the term ``cue'' is bidirectional~\cite{andrist2017looking}---referring either to (1) system-to-user cues that guide or scaffold user behavior (e.g.,~\cite{lee2022cue}), or (2) user-to-system cues that reveal the user's intent (e.g.,~\cite{rattenbury2007caad}). In this work, we focus exclusively on the latter: cues produced by users to convey their intent and interpreted by the system.}

Among these, the most common are directional cues, usually implemented via raycasting~\cite{grossman2006design, steinicke2006object}. 
Targets can be indicated by pointing with a hand~\cite{mayer2018effect, bowman1997evaluation} or controller~\cite{grossman2006design, steinicke2006object}, or by gaze~\cite{tanriverdi2000interacting, wagner2023fitts, zhang2024focusflow}.
However, when targets are small, densely arranged, or occluded, directional cues often become ambiguous for intent inference~\cite{baloup2019raycursor, lu2020investigating, yu2020fully}. 
Prior work addresses this through two main methods: immediate selection methods~\cite{bowman1997evaluation, bacim2013design, lu2020investigating} that adjust the effective selection region, and progressive refinement methods~\cite{cashion2012dense, kopper2011rapid, bacim2013design} that iteratively narrow the candidate set.
While these techniques mitigate spatial ambiguity, they remain direction-only and thus inherit the intrinsic noise of pointing; moreover, refinement-based approaches may add unnecessary overhead in simple scenes.

Therefore, a complementary line of research has explored gestural cues. Unlike direction, grasping gestures capture the geometry and functional affordances of objects~\cite{jeannerod1984timing, castiello2005neuroscience}.
VirtualGrasp~\cite{yan2018virtualgrasp} demonstrates that people naturally produce consistent grasping gestures aligned with object semantics.
However, existing approaches \cite{wobbrock2009user, piumsomboon2013user, moran2021user} often assume a one-to-one mapping between gesture and object, severely limiting scalability. 
In practice, grasping gestures support a many-to-many relationship, where a single gesture can apply to multiple semantically similar objects, increasing expressive power while introducing semantic ambiguity.

These limitations reveal a fundamental problem: \emph{no single cue can reliably support out-of-reach selection across diverse MR scenarios.} 
While combining cues seems natural, most existing multi-cue approaches~\cite{chatterjee2015gaze, yu2021gaze, lystbaek2022gaze, shi2023exploring} are rule-based---e.g., treating gaze as the primary cue and gestures as fixed command triggers~\cite{wolf2019paint}. Such deterministic designs become unreliable once the dominant cue turns ambiguous, since reliance cannot flexibly shift to alternatives.

In this work, we propose a probabilistic multi-cue integration framework for out-of-reach object selection in MR and introduce \textsc{Point\&Grasp} as its implementation, integrating directional pointing for spatial disambiguation and grasping gestures for semantic ambiguity.
Both cues are extracted from the same hand-tracking input, providing richer information without requiring additional sensors. Unlike existing deterministic multi-cue techniques, our approach treats both cues as probabilistic evidence: directional input is modeled as a likelihood over candidate objects, while a learned model estimates the gesture-object likelihood. These likelihoods are fused through Bayes’ rule to produce a posterior distribution over candidate objects, enabling robust adaptation to noisy or ambiguous cues and generalization across interaction scenarios. 

While directional cues can be readily modeled as probabilistic likelihoods, gestural cues lack such treatment. Prior methods often assume a fixed one-to-one mapping between a gesture and an object in a canonical pose, ignoring how pose and geometry affect grasp feasibility. To overcome this, we introduce a learning-based model that estimates pose-aware gesture–object likelihoods, providing context-sensitive evidence for Bayesian fusion and improving generalization across diverse interaction scenarios.

To support such modeling, we collected the \datasetfullname~(\dataset) dataset that includes both in-reach and out-of-reach grasping gestures for the same objects. Prior work has shown that in-reach gestures capture object semantics, but it remains unclear whether out-of-reach gestures convey comparable semantic information and how they differ from in-reach gestures. Our dataset addresses this gap by systematically including both positive (matched) and negative (mismatched) gesture–object pairs.
Training our probabilistic compatibility model on this dataset demonstrates robust generalization across subjects and partial transfer to previously unseen objects, providing a solid foundation for integrating grasping gestures into our probabilistic multimodal framework.

We conducted two user studies (\hyperref[sec: study_1]{Study~1} and \hyperref[sec: study_2]{Study~2}) to investigate the benefits of \textsc{Point\&Grasp} both against its constituent single cues and against state-of-the-art selection methods. 
\hyperref[sec: study_1]{Study~1} compares \textsc{Point\&Grasp} with two single-cue baselines---direction-only and grasp-only---under systematically varied spatial and semantic ambiguities.
Its goal is to identify the source of ambiguity where single-cue methods break down and how fusion overcomes these failures.
\hyperref[sec: study_2]{Study~2} then benchmarks \textsc{Point\&Grasp} against two leading selection techniques---BubbleRay~\cite{lu2020investigating} and Expand~\cite{cashion2012dense}.
These represent two distinct state-of-the-art approaches for overcoming the limitations of directional selection---adaptive assistance for immediate selection and progressive refinements, respectively.

The results reveal four central findings.
First, in \hyperref[sec: study_1]{Study~1}, \textsc{Point\allowbreak\&Grasp} consistently outperformed both single-cue baselines in selection time and completion rate, remaining robust even under the challenging spatial and semantic ambiguities.
Second, \hyperref[sec: study_1]{Study~1} showed that \textsc{Point\&Grasp} adapts cue reliance dynamically: gestural cues dominate under high spatial ambiguity, directional cues become more informative under high semantic ambiguity, and both cues converge in low-ambiguity scenarios, with fusion amplifying their combined strength. 
Third, subjective evaluations in \hyperref[sec: study_1]{Study~1} indicated that users found \textsc{Point\&Grasp} easy to learn and judged gesture-based interaction natural and consistent with everyday grasping habits.
Finally, \hyperref[sec: study_2]{Study~2} demonstrated that, when gestural cues provide reliable semantic information, \textsc{Point\&Grasp} exhibits strong robustness to ambiguity, outperforming BubbleRay in high-spatial layouts and achieving faster selections than Expand in low-spatial layouts.

Together, these findings establish probabilistic cue integration as a principled and effective solution for out-of-reach object selection. By combining the complementary strengths of directional and gestural cues, \textsc{Point\&Grasp} demonstrates how this framework improves selection robustness and flexibility, pointing toward more scalable multimodal interaction techniques for future MR systems.

In summary, the main contributions of this work are:

\begin{itemize}
    \item We propose a probabilistic cue integration framework for out-of-reach object selection in MR and implement it as \textsc{Point\&Grasp}, which fuses directional pointing and grasping gestures through Bayesian inference.
    \item We contribute \datasetfullname~(\dataset)~dataset, the first dataset for out-of-reach grasping in MR, with mid-air gestures aligned to in-reach virtual grasps as ground-truth contact references and explicit annotations of matched and mismatched gesture–object pairs.
    \item We develop a pose-aware gesture–object likelihood model based on the \dataset~dataset, which captures context-sensitive grasp feasibility, generalizes across subjects, and partially transfers to unseen objects.
    \item We validate \textsc{Point\&Grasp} through two controlled user studies, demonstrating improved performance over single-cue baselines and superior robustness compared to state-of-the-art selection techniques across diverse ambiguity conditions.

\end{itemize}

\section{Related Work}

3D out-of-reach object selection is a fundamental component of MR interaction, widely used in 3D design, games, and everyday tasks~\cite{argelaguet2013survey, zhou2022depth, ren20133d}. 
A large body of research has focused on improving its accuracy and efficiency by enabling systems to identify users’ intended targets through various interaction cues.

\subsection{Out-of-Reach Object Selection via Directional Cue}
Directional cues are widely used for out-of-reach object selection, where systems infer a target’s spatial location from a user-generated ray, most commonly via raycasting~\cite{mine1995virtual}.
The origin and direction of the ray can be defined using a handheld controller~\cite{grossman2006design, steinicke2006object, yu2019modeling, moon2024real}, the posture of a bare hand~\cite{mine1995virtual, mayer2018effect}, head~\cite{mine1995virtual, nickel2003pointing, wei2023predicting}, or gaze~\cite{nickel2003pointing, wei2023predicting}.
Some other approaches anchor the ray on specific body parts (e.g., eye, head, elbow) and extend it toward the fingertip~\cite{mine1995virtual, pierce1997image, nickel2003pointing}.
Instead of a ray, a 3D cursor allows users to specify not only a direction but also disambiguate objects at different depths, such as the Go-Go technique~\cite{zhai1994silk, poupyrev1996go}, which maps cursor direction and depth to the relative position between the user’s chest and hand.

However, directional cues are prone to errors when objects are densely arranged, small, or partially occluded~\cite{baloup2019raycursor, lu2020investigating, yu2020fully}. 
These errors stem from two types of pointing noise~\cite{mayer2015modeling, wobbrock2011effects}: constant noise, caused by geometric mismatches between the user’s sightline and the emitted ray, limited field of view, or display distortion; and random noise, arising from signal-dependent motor variability and modulated by speed–accuracy trade-offs~\cite{harris1998signal, zhai2004speed}.

To improve robustness, several immediate-selection techniques extend the basic pointing paradigm by modifying the selection endpoint or tolerance, such as cone-casting~\cite{liang1994jdcad}, or the BubbleRay~\cite{lu2020investigating}, which guarantees a unique target within an adaptive region.
While these methods help mitigate spatial ambiguity, they still rely on a single pointing action and therefore inherit user-dependent pointing noise and target-dependent biases. 
Other work adopts progressive refinement through multi-stage interaction to explicitly resolve ambiguity, including Expand~\cite{cashion2012dense}, which enlarges the selectable region as the pointer approaches the target, and SQUAD~\cite{kopper2011rapid}, which separates coarse targeting from a confirmation step.

Since interaction-level approaches cannot fully eliminate pointing noise, a complementary line of research focuses on statistical modeling of ray distributions to compensate for target-dependent offsets.
Prior work commonly models ray endpoints as Gaussian and predicts systematic deviations to improve selection accuracy~\cite{mayer2015modeling, mayer2018effect, yu2019modeling, zheng20253d}.
Building on this direction, we use distribution-based likelihoods of directional cues and integrate them with other modalities within a unified inference framework.

\subsection{Out-of-Reach Object Selection via Gestural Cue}

Gestures can serve as an alternative cue to directional information, helping disambiguate objects that spatially overlap~\cite{paulson2011object}.
Prior work has explored multiple strategies for using gestures to infer objects:
Some methods explicitly assign a unique gesture to each object~\cite{wobbrock2009user, piumsomboon2013user, moran2021user}, while others use gestures to represent object features such as shape or size.
For example, gestures that directly mimic the physical outline of objects have been investigated~\cite{pei2022hand}, emphasizing the importance of gestures being easy to discover and intuitive to learn in order to be effective~\cite{harrison2014touchtools}.

Grasping is a natural way humans interact with objects, as it reflects both the stable hand configuration afforded by an object’s shape and its functional use.
This property enables grasping gestures to convey rich information about the user’s intended object.
Prior work has shown that observing grasping gestures alone can reveal object size and shape~\cite{vatavu2013automatic}.
VirtualGrasp further demonstrated that grasping gestures can serve as a general selection cue in MR: gestures were easy to learn, and user studies showed strong cross-user agreement in how gestures mapped to objects~\cite{yan2018virtualgrasp}.
However, when multiple objects share similar shapes, grasping gestures alone may not suffice for reliable disambiguation~\cite{yan2018virtualgrasp}.

This paper addresses a research gap: how can grasping gestures, particularly performed for out-of-reach objects, be modeled probabilistically to infer user intention?
Addressing this gap requires understanding differences between virtual and physical grasping, as well as between in-reach and out-of-reach contexts, which remain underexplored.
In this work, we take the first step by constructing the \dataset~dataset for out-of-reach grasping gestures and developing a probabilistic inference model that enables grasping gestures to function as a quantitative cue for inferring user intention in out-of-reach selection tasks.

\subsection{Target Inference from Multiple Cues}

While each cue type offers unique advantages, no single cue performs reliably across all MR environments. 
This leads to our central research question: How can multiple cues be flexibly combined within a single interaction?
Despite its importance, this question has received relatively little attention.
Existing multi-cue approaches are typically limited, often assigning independent roles to each cue rather than integrating them in a principled way.
For example, gaze-hand coordination techniques often use gaze as the primary cue for object selection (e.g., pre-select a group of objects), while gestures handle fine-grained selection within densely arranged targets~\cite{chatterjee2015gaze, yu2021gaze, lystbaek2022gaze, shi2023exploring, wagner2024gaze}.
Other studies explored speech-gesture combinations, where pointing identified the object while a user’s verbal input specified its subsequent use with the object~\cite{wolf2019paint}.

While these studies demonstrate the potential of multiple cues, our work directly addresses this gap by introducing a probabilistic cue integration technique for out-of-reach selection in MR environments.
The closest approach to ours is by Wei et al.~\cite{wei2023predicting}, who predicted a user’s target of interest based on the distribution of eye endpoints and further showed that the head direction and the eye gaze distributions could be combined using Bayesian inference.
However, because these two cues provided largely overlapping rather than complementary information, their multi-cue approach did not yield significant improvements over the single-cue method based solely on gaze.
In contrast, we investigate whether integrating complementary cues with distinct discriminative power can yield measurable benefits.
Specifically, we focus on probabilistically combining pointing and grasping gestures, which address different sources of ambiguity.
Through their probabilistic integration, we aim to resolve both spatial and semantic ambiguities, thereby achieving robust out-of-reach object selection in MR.

\subsection{Datasets for Hand-Object Interaction}

Another dimension of this work concerns human-object interaction datasets, raising the question of whether existing datasets of physical, in-reach grasping can be utilized to support probabilistic modeling in out-of-reach interaction.
A variety of datasets have been introduced for physical hand–object interaction~\cite{taheri2020grab, bhatnagar2022behave, wang2022dexgraspnet, liu2022hoi4d, fan2023arctic, jian2023affordpose}.
Representative examples include GRAB~\cite{taheri2020grab}, which records full-body and hand motion during object manipulation; HOI4D~\cite{liu2022hoi4d}, which captures egocentric videos with grasping hands and objects across daily activities; and DexGraspNet~\cite{wang2022dexgraspnet}, which provides diverse dexterous grasp configurations for 3D object meshes.

These datasets have been highly valuable for studying physical in-reach grasping, supporting the training of models that predict and generate stable grasps for objects~\cite{taheri2020grab, xu2023interdiff, li2023object, zhang2024graspxl, zhang2024artigrasp} and even infer which objects are suitable for grasping~\cite{petrov2023object}.
By contrast, out-of-reach grasping involves imaginary, mid-air gestures toward distant targets.
Such gestures may differ from physical grasps due to perceptual uncertainty and noise in MR environments, yet little prior work has examined this distinction~\cite{blaga2021freehand, blaga2025vr}.
Another challenge is the need for calibration in likelihood-based inference models, which require data covering both correct (positive) and incorrect (negative) object--gesture pairings~\cite{vaicenavicius2019evaluating}.
Such balance is critical for disambiguating intent in cluttered scenes and avoiding false positives, but existing datasets overwhelmingly emphasize positive examples of feasible grasps.

To address these gaps, we contribute the \dataset~dataset of mid-air grasp gestures in out-of-reach selection scenarios, structured to support the calibrated training of probabilistic intent inference models.
To our knowledge, this is the first dataset to systematically capture out-of-reach grasping behavior, providing a foundation for robust intention modeling beyond physical grasping contexts.
\section{Probabilistic Cue Integration for Out-of-Reach 3D Object Selection}

\begin{figure*}[h]
  \centering
  \includegraphics[width=0.95\textwidth]{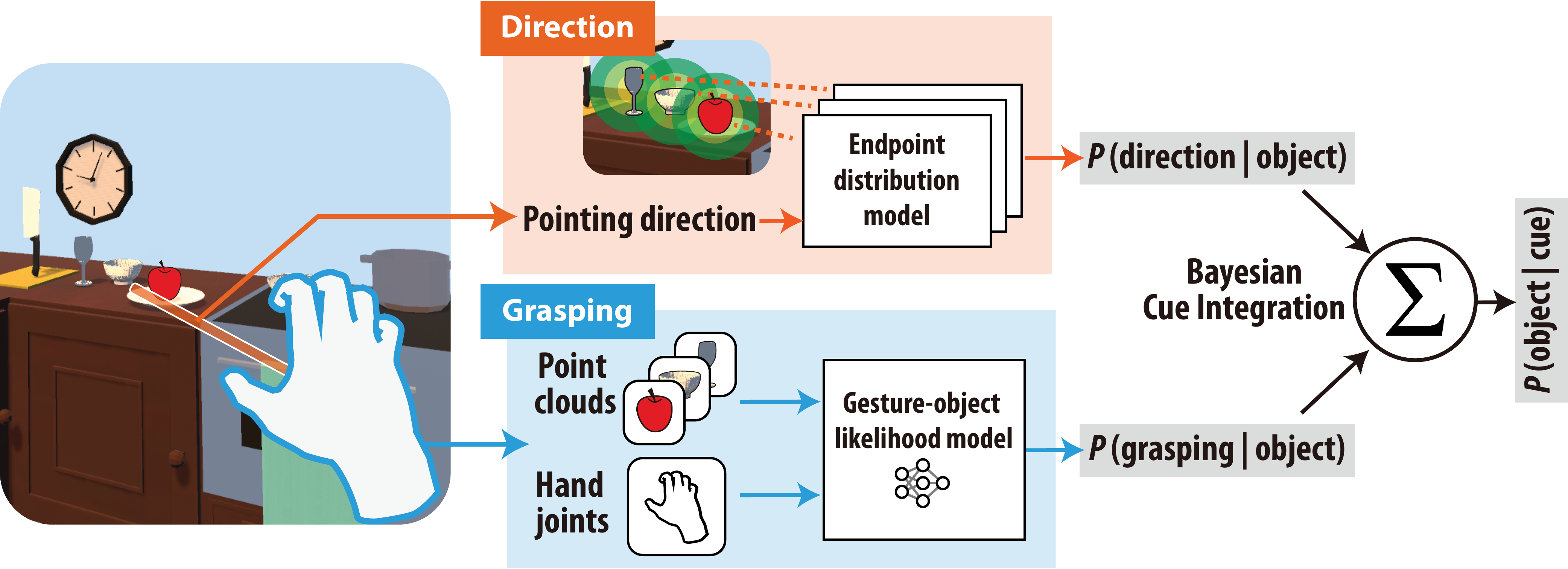}
  \caption{Overview of the proposed multi-cue inference framework for out-of-reach object selection. The framework combines an endpoint distribution model for directional cues and a gesture–object likelihood model for grasping gestural cues, which are integrated through Bayesian cue integration to infer the intended target object.}
  \Description{The figure illustrates a framework for selecting out-of-reach objects using two cues: direction and grasping. On the left, a virtual hand points toward objects on a table, including an apple, a bowl, and other items. The direction cue is shown at the top: the pointing ray connects to an “Endpoint Distribution Model” that outputs probabilities of different objects being the target. The grasping cue is shown at the bottom: the hand is represented by point clouds and joint positions, which are processed by a “Gesture–Object Likelihood Model” that estimates how well each object matches the user’s hand shape. On the right, both cues feed into a Bayesian cue integration module, depicted as a summation symbol, which combines them into a single probability distribution over objects, identifying the intended target.}
  \vspace{0.2cm}
  \label{fig:method}
\end{figure*}

Our method formulates out-of-reach 3D object selection as a probabilistic intent inference problem from multiple cues. Unlike single-cue interaction, which often fails when cues become ambiguous in specific contexts (e.g., directional cues exhibiting spatial ambiguity when objects are clustered or occluded), our framework provides a principled way to integrate heterogeneous evidence sources.

Building on this formulation, we resolve such ambiguities by casting target inference as posterior computation under a Bayesian cue-integration framework, a formulation widely studied in human perception and decision-making~\cite{ernst2002humans, lee2018moving}.

\subsection{Probabilistic Inference via Multi-Cue Integration}

We consider a set of candidate objects $\mathcal{O} = \{o_1, \dots, o_K\}$ in a given MR scene.
From a user, we consider a set of observed cues $\mathcal{C} = \{c_1, \dots, c_M\}$, where each $c_m$ corresponds to a different cue (e.g., pointing direction, hand gesture, eye gaze, speech). 
Our goal is to infer the posterior probability of each object being the intended target:
\[
p(o \mid \mathcal{C}) = p(o \mid c_1, c_2, \dots, c_M).
\]
Assuming conditional independence of cues given the target object (the naïve Bayes assumption),
\[
c_i \;\perp\; c_j \;\mid\; o \quad \forall i \neq j,
\]
the posterior distribution takes the form
\begin{equation}
\label{eq:posterior}
p(o \mid \mathcal{C}) =
\frac{p(o)\, \prod_{m=1}^M p(c_m \mid o)}
{\sum_{i=1}^K p(o_i)\, \prod_{m=1}^M p(c_m \mid o_i)}.
\end{equation}
Here, $p(o)$ represents the prior over candidate objects, and $p(c_m \mid o)$ is the likelihood of observing cue $c_m$ if the intended object were $o$.
The denominator of Eq.~\ref{eq:posterior} serves only as a normalizing constant, that is, the likelihoods play a decisive role in this inference.

Therefore, the inference procedure reduces to several steps:
For each candidate object, the system computes the likelihood of the observed cues under the assumption that the object is the intended target.
These likelihoods can be derived from previous computational models (e.g., Gaussian models of users' pointing offsets~\cite{yu2019modeling, mayer2015modeling}) that capture the statistical relationship between the intended target object and the user's resulting behavior.
The likelihoods are then combined with the prior distribution.
Finally, the object whose posterior probability is maximal is selected as the inferred target.
\[
\hat{o} = \arg\max_{o \in \mathcal{O}} \; p(o \mid \mathcal{C}),
\]

While the Bayesian framework is conceptually applicable to different cues, our work focuses on two cues that are both natural to MR interaction and complementary in their information content: \textit{directional} and \textit{gestural} cues.

\begin{itemize}
    \item \textbf{Directional cue ($c_D$):} provides spatial evidence through pointing, but can be ambiguous in scenes where objects are small, cluttered, or occluded.
    \item \textbf{Gestural cue ($c_G$):} provides object-semantic evidence by modeling the compatibility between the user’s hand configuration and object shape, but may be ambiguous among objects of similar shape or category.
\end{itemize}
To compute the posterior $p(o \mid c_D)$ and $p(o \mid c_G)$, we must therefore model $p(c_D \mid o)$ and $p(c_G \mid o)$.

\subsection {Directional Cue Inference}
The directional cue provides spatial evidence for inferring the intended target.
As illustrated in Figure~\ref{fig:method}, we represent directional intent with a ray defined by an origin and a direction vector. The ray is cast from the user’s hand toward the scene and intersects the reference plane of the objects at an endpoint $p_e$.  

Following prior work~\cite{yu2019modeling, mayer2015modeling}, we model the directional cue likelihood $p(c_D \mid o)$ as a Gaussian likelihood:
\[
p(c_D \mid o) \propto 
\exp\!\left(-\frac{\left(p_e - p_o\right)^2}{2\sigma_d^2}\right),
\]
Here $p_o$ denotes the center of candidate object $o$, serving as the Gaussian mean, and $\sigma_d$ controls the sensitivity to spatial deviation.  
This definition reflects the probability that the observed endpoint $p_e$ would result if the user were actually intending to select object $o$. Consequently, objects whose centers are closer to the ray endpoint receive higher likelihood values, meaning the directional cue provides stronger support for them as the intended target.

\subsection{Grasping Gestural Cue Inference}

The grasping gestural cue captures semantic information about the object, reflecting its shape, size, and affordances. In other words, a grasping hand posture carries cues about which objects in the scene are physically compatible candidates, even without relying on precise directional pointing.

To make use of this information, we formulate a gesture–object likelihood model $p(c_G \mid o)$ that estimates the probability of a gesture $c_G$ being intended for a candidate object $o$:
\[
p(c_G \mid o) \approx f_\theta(c_G, o),
\]
where $f_\theta$ is parameterized by a neural network that takes as input (1) the user’s hand configuration (joint positions or gesture embedding) and (2) object features such as shape descriptors or category embeddings. The output is normalized to form a probability distribution over candidate objects.
The choice of model architecture and the training objective for $f_\theta$ depends on available input data representations.
For example, architectures may range from simple MLPs to graph- or attention-based models, depending on how hand and object features can be acquired in practice.
Likewise, one may use contrastive objectives when the training dataset provides positive and negative gesture–object pairs, or a standard cross-entropy loss when only ground-truth target object labels are provided.
Note that the specific architecture and training procedure used in this work are described in Section~\ref{sec:gesture cue}.

\subsection{Implementation of \textsc{Point\&Grasp}}
Figure~\ref{fig:method} provides an overview of how the \textsc{Point\&Grasp} method is instantiated in our system. The general Bayesian formulation described earlier requires us to specify likelihood models for both directional and gestural cues. In our VR implementation, this involves defining how the pointing ray is constructed, how hand postures and objects are represented, and how these representations are processed to produce cue likelihoods that are ultimately combined in the final inference.

\subsubsection{Directional Cue $c_D$} \label{sec: directional_cue}


The directional cue fundamentally depends on how the ray is defined. Prior work~\cite{lee2003evaluation, argelaguet2013survey} has shown that head–hand direction is a common basis for defining pointing, where the ray originates at the user’s head and passes through a hand-based endpoint. While some techniques~\cite{schwind2018up} use the fingertip as this endpoint, our setting requires a reference point that remains consistent across different grasp poses. We therefore adopt the wrist position as the ray endpoint, which preserves the head–hand relationship while providing a stable spatial reference during grasping.

In our model, $\sigma_d$ specifies the size of the region around the object's center where the directional cue retains meaningful probability.
We use a constant $\sigma_d$ to avoid unintentional differences in selection difficulty across objects, assuming graspable everyday objects whose visual sizes fall within a reasonably limited range.
Based on prior observations in~\citet{yu2019modeling}, we adopt $\sigma_d = 0.4$ as a representative value for objects with a visual angle of $1^\circ$ to $2^\circ$.
Note that size-dependent formulations (i.e., increasing $\sigma_d$ according to the object size) can be considered in other settings.

\subsubsection{Gestural Cue $c_G$} \label{sec:gesture cue}

To effectively model the likelihood of gestural cues given an object, we require a representation that captures both the object and hand configuration.
We represent the hand through its 3D joint positions and the objects through its point cloud.
The point cloud is defined in the canonical coordinate system, where the object center is the object's origin and the canonical orientation defines the axes. 
The point cloud directly reflects the object’s shape and pose; i.e., different orientations of the same object are expressed as different point coordinate sets. 
For the hand, we represent joint positions relative to the wrist as the origin, while using the same canonical axes as those of the object’s point cloud. 

In this way, the representation captures the relative geometry between hand and object: regardless of where the object is located, how it is oriented, or how far the hand is from it, the combination of point cloud and wrist-centered hand joints provides a consistent description of the object’s shape and the hand–object relationship.

The gesture-object likelihood model comprises two encoders: one for the hand and one for the object.
\begin{itemize}
\item \textbf{Hand encoder}. The hand encoder takes the transformed joint positions (63-dimensional input) and processes them through a multi-layer perceptron (MLP). The encoder consists of two fully connected layers with 256 hidden units each. Each layer is followed by batch normalization and ReLU activation, with a dropout layer ($p=0.2$) between the two blocks. The output is a 256-dimensional gesture embedding.
\item \textbf{Object encoder}. The object encoder first converts the raw point cloud into a basis point set (BPS) representation~\cite{prokudin2019efficient}, yielding a compact 1024-dimensional vector that captures the object’s geometry independently of the number of points. This vector is then passed through an MLP with two fully connected layers (1024$\rightarrow$512 and 512$\rightarrow$256), each followed by batch normalization, ReLU, and dropout ($p=0.2$). The output is a 256-dimensional object embedding, aligned in dimensionality with the hand embedding.
\end{itemize}
The gesture and object embeddings are then concatenated (512-dimensional vector) and fed into a fusion MLP. This consists of two hidden layers (512$\rightarrow$128 and 128$\rightarrow$64), each followed by batch normalization, ReLU, and dropout ($p=0.2$). Finally, a linear layer maps to a single scalar output, which is squashed by a sigmoid activation to produce the probability $p(c_G \mid o)$ that the observed gesture corresponds to the candidate object.

The model is trained with a binary cross-entropy loss on each gesture–object pair, where $y \in \{0,1\}$ indicates whether the gesture is intended for the object:
\[
\mathcal{L} = - \big[ y \log p(c_G \mid o) + (1 - y) \log (1 - p(c_G \mid o)) \big],
\]

\subsubsection{Combining Cues} 
In practice, the \textsc{Point\&Grasp} technique requires the user to simultaneously provide both types of evidence: 
\begin{itemize}
    \item By extending the arm and aligning the wrist with the target, the user produces a ray that provides the directional cue $c_D$. 
    \item By shaping the hand into a natural grasp posture, the user conveys semantic information about the intended object, forming the gestural cue $c_G$. 
\end{itemize}

During inference, the system evaluates for each candidate object $o$ its directional likelihood $p(c_D \mid o)$ and its gestural likelihood $p(c_G \mid o)$. These two terms are then combined under the Bayesian formulation as Eq.~\ref{eq:posterior}. We assume $p(o)$ as uniform prior over candidate objects in the 3D scenes, meaning that each object is considered equally likely to be the target. 
This prior can also be flexibly adjusted; for example, to encode biases such as certain objects being more frequently used.
\section{\datasetfullname~Dataset and Model Evaluation}

\begin{figure*}[t]
  \centering
  \includegraphics[width=0.8\textwidth]{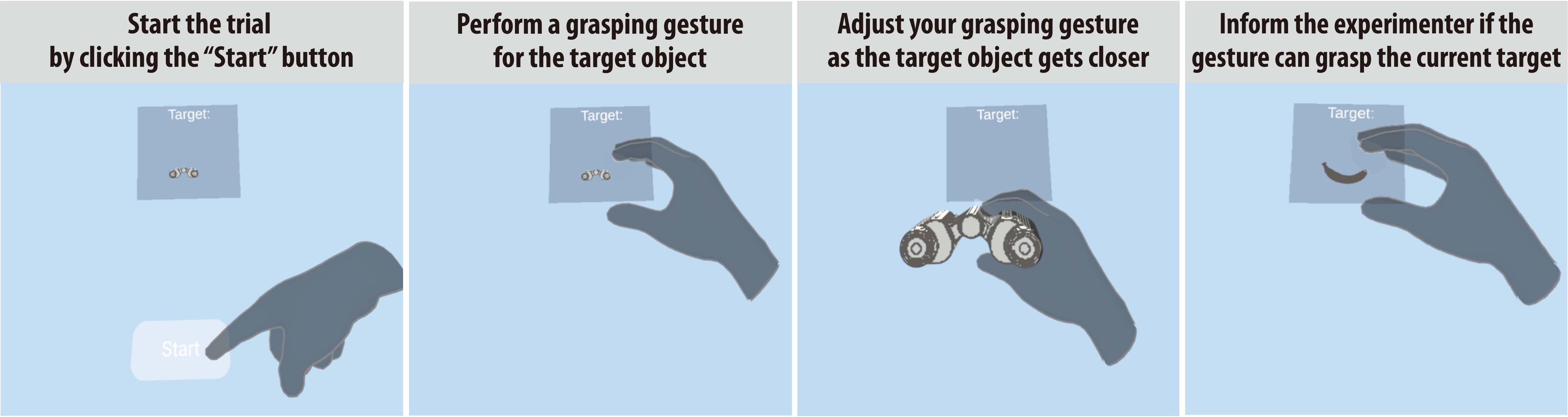}
  \caption{The trial procedure in the VR data collection task. (1) Participants initiated the trial by pressing a virtual “Start” button and performed the natural grasping gesture they would normally use to pick up the object. (2) The object was then placed within reach, and participants indicated the intended grasp location before refining their posture to align with the object. (3) The finalized grasping posture was tested for compatibility with five additional objects, for which participants verbally indicated whether their current gesture could plausibly grasp the object.}
  \Description{The figure illustrates the step-by-step procedure of the VR data collection task across four panels. In the first panel, participants begin by pressing a virtual “Start” button. In the second panel, a target object appears on screen and participants perform a natural grasping gesture they would normally use to pick it up. In the third panel, the target object is shown closer, and participants adjust their grasping gesture to better align with the object’s shape and position. In the fourth panel, participants confirm to the experimenter whether their current gesture could plausibly grasp the presented target.}
  \vspace{0.2cm}
  \label{fig:data_collection}
\end{figure*}

Grasping gestures encode both the geometric properties of objects and their functional affordances, making them a powerful cue for inferring user intent. However, existing datasets such as GRAB~\cite{taheri2020grab} primarily focus on in-reach physical grasps and do not capture how grasping gestures are produced when objects are out of reach, a setting that is central to mid-air interaction in MR.

To address this gap, we introduce the \datasetfullname~(\dataset)~dataset, which captures both in-reach and out-of-reach grasping gestures. In-reach gestures provide ground-truth contact information, while their out-of-reach counterparts reveal users’ intended grasps without physical interaction. We analyze the dataset to characterize the distribution, variability, and semantic structure of out-of-reach gestures, including their differences from in-reach gestures and their gesture-induced object confusability.

To ensure that our interaction technique receives stable and reliable probability $p(c_G \mid o)$ during runtime, we evaluate several network architectures. Because gesture and object features have distinct characteristics, we adopt a dual-branch encoder (Section~\ref{sec:gesture cue}) to extract their representations separately before fusing them. The rationale and effectiveness of this design are validated in Section~\ref{sec:model training and evaluation}.

\subsection{Method: Data Collection}
\subsubsection{Participants}
19 participants (13 male, 6 female; aged 24–37) were recruited to participate in data collection. All participants were right-handed and had no motor or vision impairments based on self-report. 
Before the experiment, each participant received a detailed briefing and provided informed consent.

\subsubsection{Experimental Design}
Each \emph{trial} was designed to sequentially collect three types of data for a given object from each participant in the VR environment: (1) an out-of-reach grasping gesture, (2) an in-reach grasping gesture, which also served as a ground-truth contact reference, and (3) subjective evaluations of gesture–object compatibility with other objects of varying shapes. This design yields annotations of both positive and negative samples under out-of-reach and in-reach conditions.

We employed a \emph{within-subjects design} with \emph{target object} as the control factor, in which each participant completed trials with the same 30 target objects presented in a randomized order.
To maximize the diversity of grasping behaviors, participants performed three trials per object, each with a distinct object pose. For each pose, they were asked to produce a grasping gesture they would normally use to grasp or interact with the object, from the perceived object shape and orientation.

Across the three trials, only object orientation varied. Each object was assigned a canonical orientation reflecting its typical usage (e.g., an upright mug, a horizontal knife).
To introduce noticeable variation across trials while maintaining realism, we first sampled a base yaw angle from $\{0^\circ, 120^\circ, 240^\circ\}$, and applied a random perturbation within $\pm 60^\circ$, ensuring both systematic coverage of diverse grasping orientations and natural variability across repeated presentations.

The 30 target objects were common household items widely used in hand–object interaction research~\cite{taheri2020grab, cho2024dense}, spanning diverse shapes and affordances to elicit varied grasping strategies.
Each participant completed 90 trials (30 objects × 3 orientations), totaling over 1,700 trials across all users—providing broad coverage of gesture–object pairings while keeping each session under one hour.

\subsubsection{Task}
Each trial followed a sequence of three stages (see Figure~\ref{fig:data_collection}).
In the first stage, the target object appeared floating at eye level, 2 meters in front of the participant. The trial began when participants pressed a start button with their dominant hand, after which they were asked to perform the \emph{``natural grasping gesture you would normally use if you were to pick up the object.''}
Because the object was intentionally placed out of reach, participants could not physically interact with it and therefore relied solely on prior experience to imagine the grasp. Upon completion, they verbally notified the experimenter.
This out-of-reach stage enabled us to capture grasping intention without influence from physical contact.

In the second stage, the same target object was moved into the participant’s reachable space (approximately at arm’s length) while maintaining its orientation. Participants were asked to bring their hand to the spatial location they had previously imagined and then refine the posture as needed to naturally grasp the now reachable object.
This stage recorded the in-reach gesture together with its precise spatial relationship to the object. Once the gesture was finalized, participants notified the experimenter.

In the third stage, participants maintained their finalized grasp while the target object was sequentially replaced with five objects randomly selected from the remaining pool. For each new object, they verbally indicated whether the current hand posture could plausibly grasp it (“yes,” “no,” or “unsure”), and the experimenter recorded the response.
If the response was “yes,” the object was moved into reachable space as in stage two, and participants reassessed whether the grasp was feasible. When the answer again was “yes,” we recorded the grasp and its spatial configuration; otherwise, only the compatibility labels were stored.

\subsubsection{Study Procedure}
Before the experiment, participants provided written informed consent and were given time to familiarize themselves with the VR environment through practice trials. We also measured each participant’s arm length to define the boundary of their reachable space and ensure correct object placement during in-reach stages.
During data collection, participants progressed through the trials at their own pace and could rest between trials as needed. On average, each session (90 trials) lasted about one hour.

\begin{figure*}[t]
    \centering
    \includegraphics[width=0.9\linewidth]{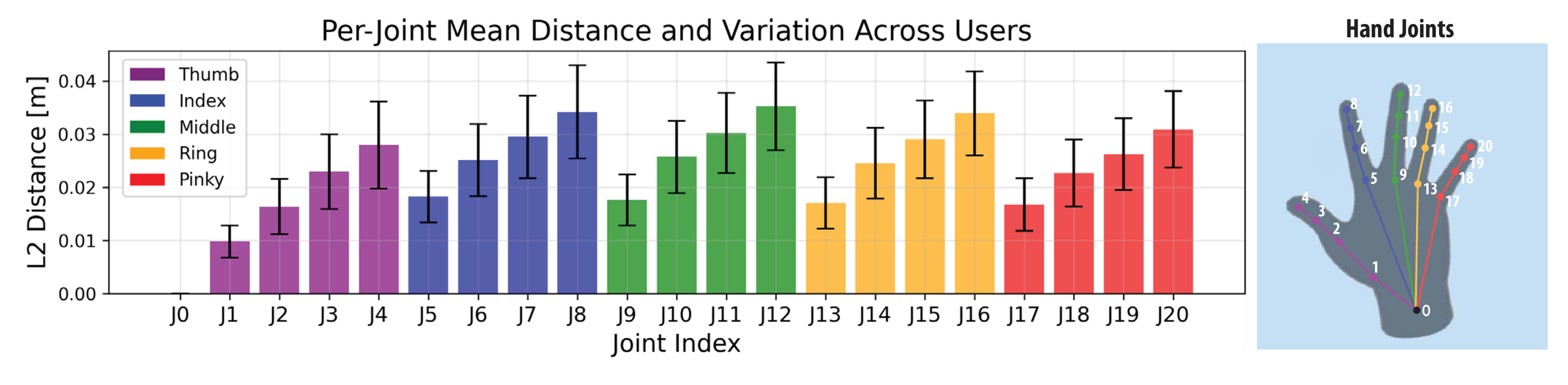}
    \caption{The results show that hand gesture for an out-of-reach object is significantly different from when it is within reach, and that variability across users is high. Bars indicate per-joint mean L2 distance grouped by finger, with error bars showing standard deviations. The illustration maps each joint index (J0–J20) to its anatomical position on the hand for reference.}
    \Description{The figure shows a bar chart of per-joint mean L2 distance and variation across users for in-reach versus out-of-reach gestures, alongside a reference illustration of a hand with labeled joints. The x-axis lists joint indices (J0–J20), grouped by finger: thumb (purple), index (blue), middle (green), ring (yellow), and pinky (red). The y-axis indicates L2 distance in meters. Each bar represents the average L2 distance for a joint, with error bars showing standard deviations across users. The chart shows that out-of-reach gestures differ substantially from in-reach ones, with larger distances and high variability across joints. On the right, a hand diagram maps each joint index (0–20) to its anatomical position for reference.}
    \label{fig:joint_mean_distance}
\end{figure*}

\begin{figure}[h!]
    \centering
    \includegraphics[width=0.9\linewidth]{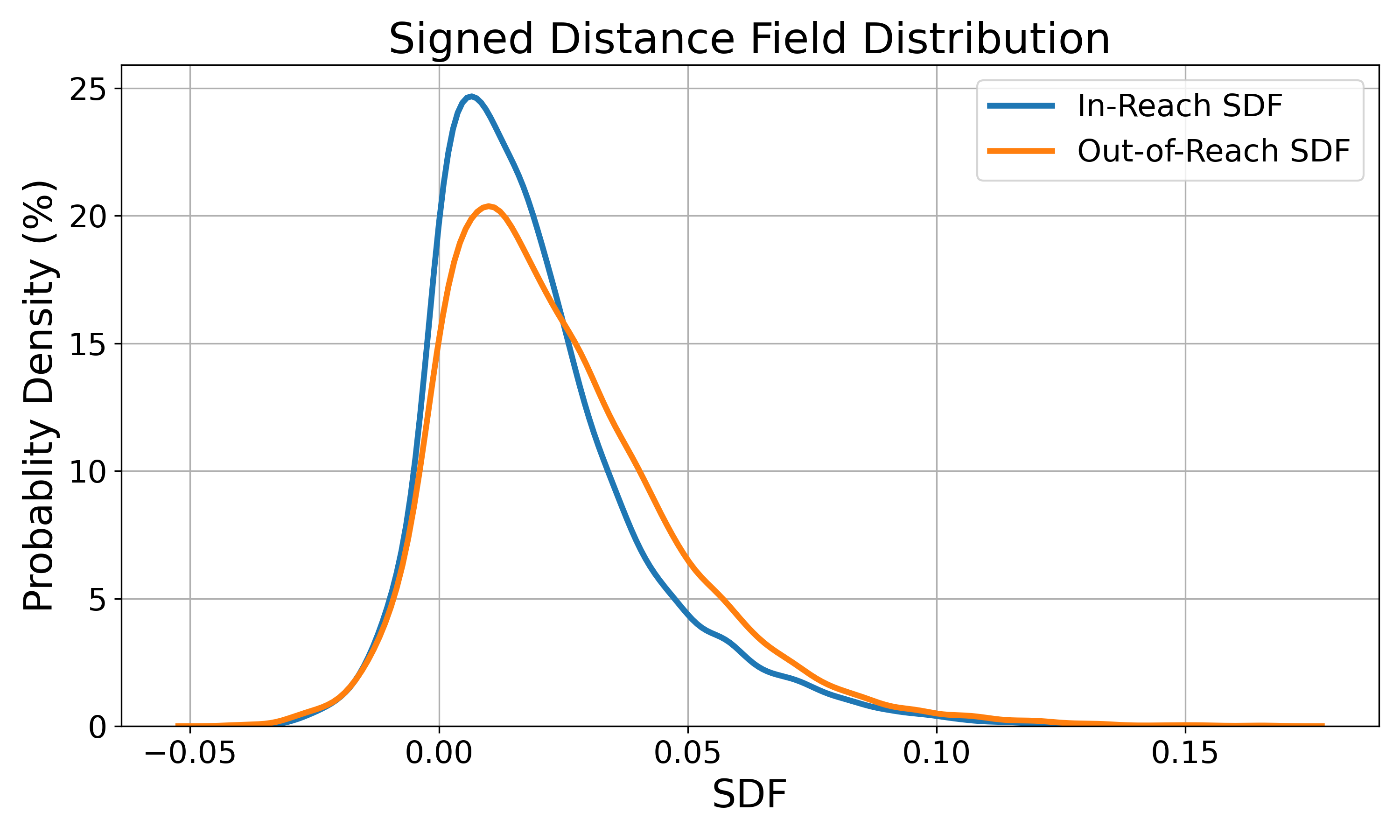}
    \caption{Signed Distance Field (SDF) distributions for in-reach and out-of-reach gestures. Out-of-reach gestures show a rightward shift, indicating more extended hand postures.}
    \Description{The figure presents a line graph comparing Signed Distance Field (SDF) distributions for in-reach and out-of-reach hand gestures. The x-axis represents SDF values, and the y-axis shows probability density as a percentage. Two curves are shown: a blue line for in-reach gestures and an orange line for out-of-reach gestures. Both curves have a similar bell-shaped profile, but the orange curve is shifted slightly to the right, indicating that out-of-reach gestures involve more extended hand postures.}
    \label{fig:sdf_distribution}
\end{figure}

\subsubsection{Apparatus and Materials}
The experiment was conducted in a virtual environment developed with Unity3D and displayed on a Meta Quest 2 headset with its native hand tracking.
The headset operated in a wired configuration connected to a desktop computer, ensuring stable rendering at 90 Hz and reliable data transfer throughout the experiment.

\subsection{\datasetfullname~Dataset Analysis}
\subsubsection{Dataset Statistics}
The \dataset~dataset comprises 10,260 gesture–object pairs, each annotated with binary compatibility labels for both in-reach and out-of-reach conditions. Most pairs were judged incompatible in both conditions (68.89\%, 7,068 samples), while clearly compatible cases account for 23.62\% (2,423 samples). The remaining samples reflect asymmetric or uncertain judgments: 3.69\% (379) were incompatible in-reach but compatible out-of-reach, 2.23\% (229) incompatible in-reach but uncertain out-of-reach, and 1.57\% (161) uncertain in-reach but compatible out-of-reach. Overall, the distribution contains a large proportion of clear positives and negatives, while also capturing natural ambiguity that is valuable for training probabilistic intent models.

\subsubsection{Differences Between In-Reach and Out-of-Reach Gestures}
\label{sec:gesture_analsysis}
To assess whether out-of-reach gestures are semantically comparable to in-reach gestures while revealing systematic differences, we compared their Signed Distance Field (SDF) distributions. SDF characterizes spatial occupancy by measuring the signed distance from a point to the nearest object surface, with positive values outside the object, negative values indicating penetration, and near-zero values reflecting surface contact~\cite{osher2004level, frisken2000adaptively}.
A two-sample Kolmogorov–Smirnov test~\cite{massey1951kolmogorov} revealed a significant difference between in-reach and out-of-reach distributions ($p < 0.001$). As shown in Figure~\ref{fig:sdf_distribution}, out-of-reach gestures exhibit a slight rightward shift, indicating more extended hand postures at greater distances.

To further localize these differences, we computed per-joint Euclidean distances across conditions. As illustrated in Figure~\ref{fig:joint_mean_distance}, most joints differ by 1.5–3 cm on average, with the largest deviations at distal joints (fingertips) and smaller differences at proximal joints. Participant variation remained limited, indicating a consistent deviation pattern across users.

\begin{figure*}[h]
    \centering
    \includegraphics[width=0.8\linewidth]{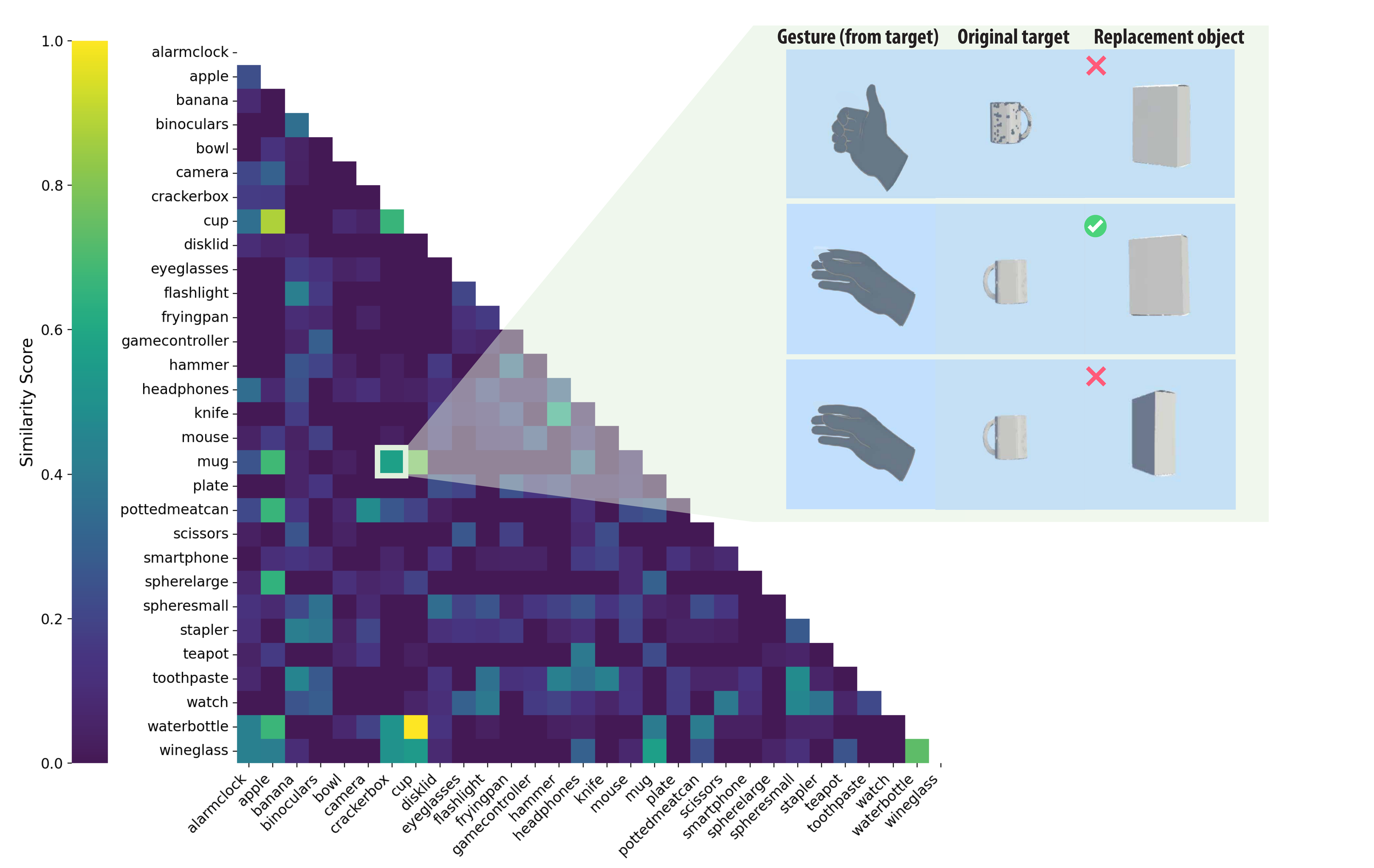}
    \caption{The object confusion matrix shows that when using grasping gesture cues alone, some objects are more easily and some less easily confused with each other. 
    Each cell indicates how often a gesture elicited by the target object was also judged compatible with a replacement object; brighter values denote higher confusion likelihood.
    The examples on the right illustrate typical outcomes: (Row 1) the gesture for a mug was incompatible with a crackerbox; (Row 2) another gesture for a mug was judged compatible with a crackerbox; (Row 3) although the same gesture for a mug as in Row 2 was applied, the replacement object could not be grasped in the current hand pose due to its orientation. These cases highlight the many-to-many nature of gesture–object relationships, where one gesture can be judged compatible with multiple objects, and the same object can be grasped with different gestures depending on its orientation.}
    \Description{The figure presents an object confusion matrix showing how often grasping gestures for one object are judged compatible with another. Along both axes, object categories are listed (e.g., apple, mug, scissors, smartphone, teapot). Each cell represents the similarity score between a target object’s gesture and a possible replacement object, with brighter colors indicating higher likelihood of confusion. The diagonal cells are generally darker, reflecting correct matches, while off-diagonal bright spots show confusion cases.
    On the right, three example cases are illustrated. In the first row, a gesture for a mug was judged incompatible with a crackerbox. In the second row, another mug gesture was judged compatible with a crackerbox (highlighted with a green checkmark). In the third row, the same gesture for a mug as in row two was judged incompatible because the replacement object could not be grasped in the current hand orientation. These examples emphasize the many-to-many nature of gesture–object relationships: a single gesture can match multiple objects, and the same object may be judged differently depending on gesture orientation.
    }
    \label{fig:object_confusion_matrix}
\end{figure*}

\subsubsection{Gesture-Induced Object Confusability}
To examine the semantic discriminability of out-of-reach gestures, we constructed an object–object confusion matrix based on the compatibility judgments collected in Stage 3 of the experiment (Figure~\ref{fig:object_confusion_matrix}). In this stage, participants maintained a given hand posture while the target object was sequentially replaced with other objects, and they indicated whether the posture remained compatible. Aggregating these judgments across participants and trials yields an average similarity score for each object pair.

Overall, most objects are clearly separable in gesture space, suggesting that participants produced distinct hand postures reflecting object function and shape. However, certain object pairs exhibit high confusability, such as cup–mug–wineglass, which frequently share similar hand configurations. This pattern indicates a many-to-many relationship between gestures and objects, where a single gesture can be compatible with multiple objects and a single object can elicit diverse gestures. These findings delineate the semantic boundaries of out-of-reach gestures and motivate the inclusion of both positive and negative samples for probabilistic inference.

\begin{figure}
    \centering
    \includegraphics[width=\linewidth]{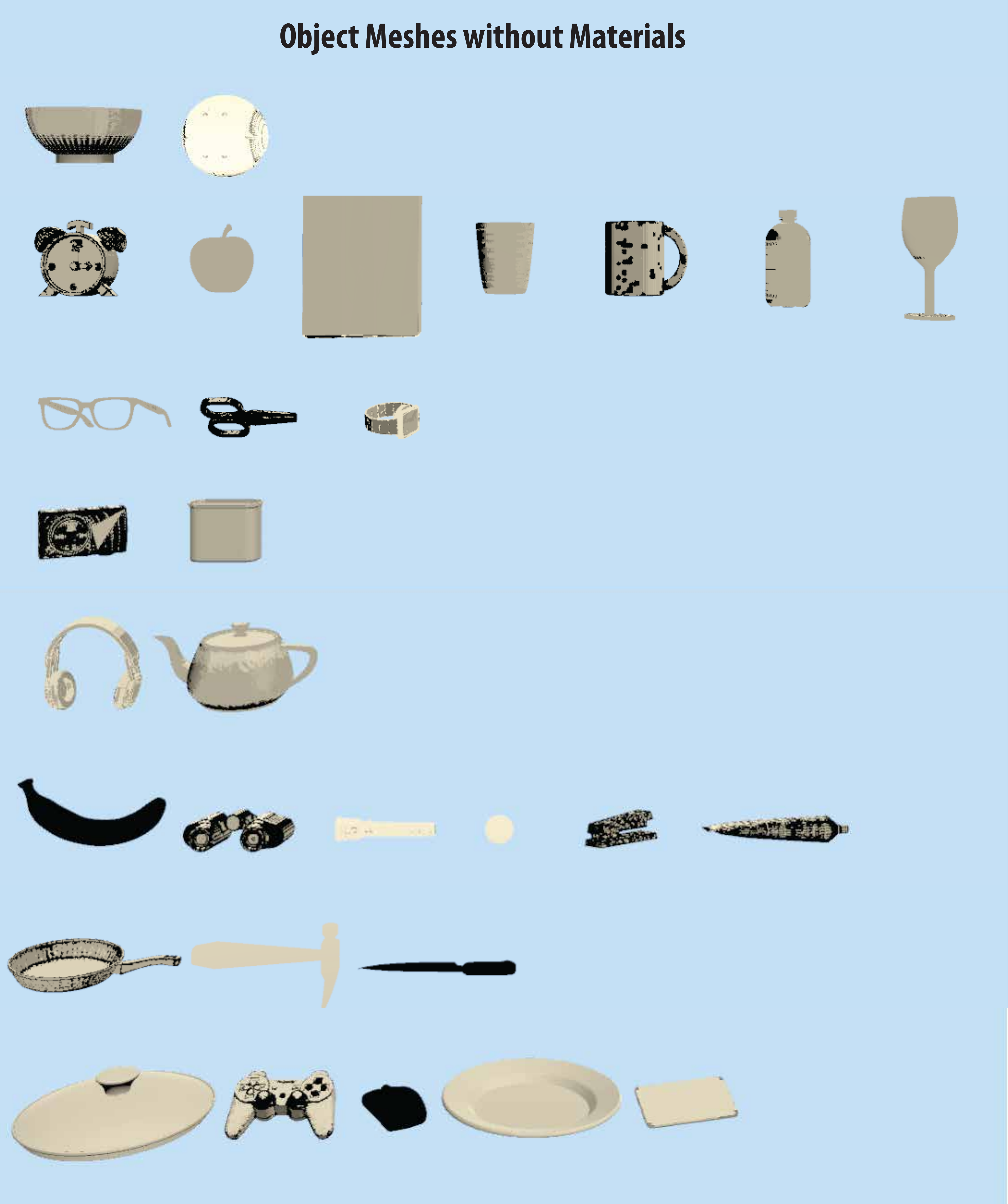}
    \caption{Clustering analysis shows that gesture-based confusability is systematic rather than random: objects that afford similar grasping postures, such as fryingpan–hammer–knife or cup–mug–wineglass–waterbottle, naturally group together. The clusters were derived by applying the Silhouette method to the similarity matrix of 30 objects: Cluster 1: bowl, large sphere. Cluster 2: alarm clock, crackerbox, apple, mug, water bottle, wineglass. Cluster 3: eyeglasses, scissors, watch. Cluster 4: camera, potted meat can. Cluster 5: headphones, teapot. Cluster 6: banana, binoculars, flashlight, small sphere, stapler, toothpaste. Cluster 7: frying pan, hammer, knife. Cluster 8: disklid, game controller, mouse, plate, smartphone.}
    \Description{The figure displays 30 everyday object meshes without textures, arranged in rows against a light blue background. Each row corresponds to one cluster derived from the similarity matrix, grouping objects that afford similar grasping postures.
    Row 1 (cluster 1): Bowl, large sphere.
    Row 2 (cluster 2): Alarmclock, crackerbox, apple, mug, waterbottle, wineglass.
    Row 3 (cluster 3): Eyeglasses, scissors, watch.
    Row 4 (cluster 4): Camera, potted meat can.
    Row 5 (cluster 5): Headphones, teapot.
    Row 6 (cluster 6): Banana, binoculars, flashlight, small sphere, stapler, toothpaste.
    Row 7 (cluster 7): Frying pan, hammer, knife.
    Row 8 (cluster 8): disklid, gamecontroller, mouse, plate, smartphone.
    The clustering shows that confusable objects group systematically rather than randomly — for instance, frying pan–hammer–knife (elongated grasp), or cup–mug–wineglass–water bottle (cylindrical grasp).}
    \label{fig:clusters}
\end{figure}

To further analyze the structure of this confusability, we applied the Silhouette method to the similarity matrix and clustered the 30 objects (Figure~\ref{fig:clusters}). The resulting clusters reveal systematic gesture-induced groupings: for example, fryingpan–hammer–knife form a coherent cluster associated with tool-like grasping postures, while cup–mug–wineglass–waterbottle cluster together due to shared cylindrical affordances. These results show that confusability is structured rather than random, reflecting functional and morphological regularities in gesture–object mappings.

\subsection{Model Training and Evaluation} \label{sec:model training and evaluation}
The goal of model evaluation is to assess the stability and suitability of the
gesture–object likelihood model that supports our interaction technique \textsc{Point\&Grasp}. During interaction, the model continuously estimates the probability $p(c_G \mid o)$, and therefore must produce consistent and reliable outputs across different users and different objects. To assess this, we evaluate the model under three complementary generalization settings:

\begin{itemize}
    \item Within-subject evaluation: training and testing on disjoint trials from the same participants. 
    \item Cross-subject evaluation: training on a subset of participants and testing on unseen participants. 
    \item Cross-object evaluation: training on a subset of objects and testing on unseen objects. 
\end{itemize}

Given the substantial structural differences between gesture and object features, we adopt a dual-branch encoder to extract their representations separately before fusing them (see Section~\ref{sec:gesture cue} for details). To validate the effectiveness of this design for our task, we compare it against a simplified single-branch alternative. These comparisons confirm the benefits of the dual-branch architecture and provide a reusable baseline for future research on gesture–object compatibility tasks.

\subsubsection{Training Setup}

We generated three train/test splits corresponding to the three generalization protocols described above.

\begin{itemize}
    \item Within-subject split: each participant’s trials were divided into an 8:2 train/test ratio.
    \item Cross-subject split: the \dataset~dataset is split by subjects, following a 7:3 train/test ratio.
    \item Cross-object split: objects were partitioned according to clustering analysis shown in Figure~\ref{fig:clusters}, with one representative object from each cluster held out for testing. Models were not exposed to these objects during training. 
\end{itemize}

Implementation details of the single-branch baseline and the training configuration are provided in the supplementary material.

\subsubsection{Evaluation Results}
Table~\ref{tab:training_result} summarizes the model performance across the three generalization settings. Overall, the dual-branch model outperforms the single-branch baseline on most metrics, indicating that separately encoding gesture and object features provides a clear advantage for this task.
In the within-subject setting, both models achieve strong performance with only minor differences, suggesting that both architectures are able to capture individual users’ compatibility judgment patterns when the training and testing data come from the same participant.
In the cross-subject setting, the dual-branch model achieves higher accuracy, recall, and precision, while the single-branch baseline yields a slightly lower ECE. These results show that separating gesture and object encoding leads to more stable representations that generalize better to unseen users.
The cross-object setting poses the greatest challenge, with performance decreasing for both models due to the difficulty of inferring compatibility for entirely unseen object shapes and affordances. Nevertheless, the dual-branch model consistently outperforms the single-branch baseline across all metrics, demonstrating stronger shape-level generalization.
Taken together, these results validate the effectiveness of the dual-branch architecture, particularly under cross-user and cross-object distribution shifts, while also revealing room for improvement when generalizing to completely novel objects.

\begin{table}[h]
    \centering
    \setlength{\tabcolsep}{6pt}
    \renewcommand{\arraystretch}{1.15}
    \caption{Evaluation results across three settings. 
    Accuracy, Recall, and Precision ($\uparrow$) are higher-is-better; 
    ECE ($\downarrow$) is lower-is-better.}
    \begin{tabular}{lrrrr}
    \specialrule{1.2pt}{0pt}{2pt}   
    Dataset 
      & \multicolumn{1}{c}{Accuracy $\uparrow$}
      & \multicolumn{1}{c}{Recall $\uparrow$}
      & \multicolumn{1}{c}{Precision $\uparrow$}
      & \multicolumn{1}{c}{ECE $\downarrow$} \\
    \midrule
    \multicolumn{5}{l}{\textbf{Within-subject}} \\
    Single-branch        & 0.813 & \textbf{0.837} & 0.791 & \textbf{0.070} \\
    \textbf{Ours} & \textbf{0.816} & 0.807 & \textbf{0.821} & 0.085 \\
    \midrule
    \multicolumn{5}{l}{\textbf{Cross-subject}} \\
    Single-branch        & 0.811 & 0.835 & 0.763 & \textbf{0.070} \\
    \textbf{Ours} & \textbf{0.822} & \textbf{0.843} & \textbf{0.775} & 0.090 \\
    \midrule
    \multicolumn{5}{l}{\textbf{Cross-object}} \\
    Single-branch        & 0.529 & 0.296 & 0.534 & 0.338 \\
    \textbf{Ours} & \textbf{0.566} & \textbf{0.368} & \textbf{0.590} & \textbf{0.299} \\
    \specialrule{1.2pt}{2pt}{0pt}   
    \end{tabular}
    \label{tab:training_result}
\end{table}

\section{Overview of User Studies}

As discussed earlier, single-cue techniques for out-of-reach object selection have inherent limitations. Directional cues are intuitive but degrade under spatial ambiguity, such as when targets are densely clustered or partially occluded, leading a single pointing ray to intersect multiple candidates. In contrast, grasping gestural cues convey object semantics but become less discriminative under semantic ambiguity, when objects share similar shapes or functions and appear equally compatible with the same gesture.

To investigate how these limitations can be mitigated, we evaluated \textsc{Point\&Grasp} through two complementary user studies. 
The first examines its behavior and performance relative to single-cue baselines (\hyperref[sec: study_1]{Study~1}), while the second compares our method against state-of-the-art 3D selection techniques (\hyperref[sec: study_2]{Study~2}). 
Together, the studies address the following research questions:

\begin{itemize}
    \item \textbf{RQ1:} Does the Interaction method \textsc{Point\&Grasp} that integrates multiple cues outperform single-cue baselines (\textsc{Point} or \textsc{Grasp}) across various selection contexts?
    \item \textbf{RQ2:} How do users exploit the flexibility afforded by \textsc{Point\&\-Grasp} across selection contexts? Which cues are used when resolving spatial or semantic ambiguity? 
    \item \textbf{RQ3:} How do users subjectively perceive \textsc{Point\&Grasp}?
    \item \textbf{RQ4:} How does \textsc{Point\&Grasp} compare with state-of-the-art 3D selection techniques?
\end{itemize}

\section{Study 1: Comparing with Single-Cue Baselines} \label{sec: study_1}
We first examine how each single cue contributes to hand-based selection performance and what advantages arise from probabilistically integrating them.
To this end, we compare \textsc{Point\&Grasp} with its two single-cue variants---\textsc{Point} and \textsc{Grasp}---to assess accuracy, cue usage patterns, and subjective user preferences.
\vspace{1em}

\subsection{Method}

\subsubsection{Apparatus}
The study was conducted using a Meta Quest 2 headset connected via Oculus Link, with participants remaining seated throughout the experiment. The right hand was used for directional input and grasping gesture, while the left hand confirmed selections via the space bar on a keyboard.
A custom prototype was implemented in Unity and communicated in real time with a Python-based gesture–object likelihood model.
Gesture data were transmitted from Unity to Python, where likelihoods were computed and returned to Unity. 
The system was executed on a desktop computer with an Intel Core i7-4790K CPU (4.00 GHz), 16 GB RAM.

\subsubsection{Participants}
Twelve participants (8 male, 4 female; aged 26–30) were recruited from the university community. All were right-handed with normal or corrected-to-normal vision, hearing, and motor abilities. Their self-reported VR experience was moderate ($M = 2.75$ on a 5-point scale, 1 = None, 5 = Expert). Participation was voluntary with informed consent.

\begin{figure}
    \centering
    \includegraphics[width=\linewidth]{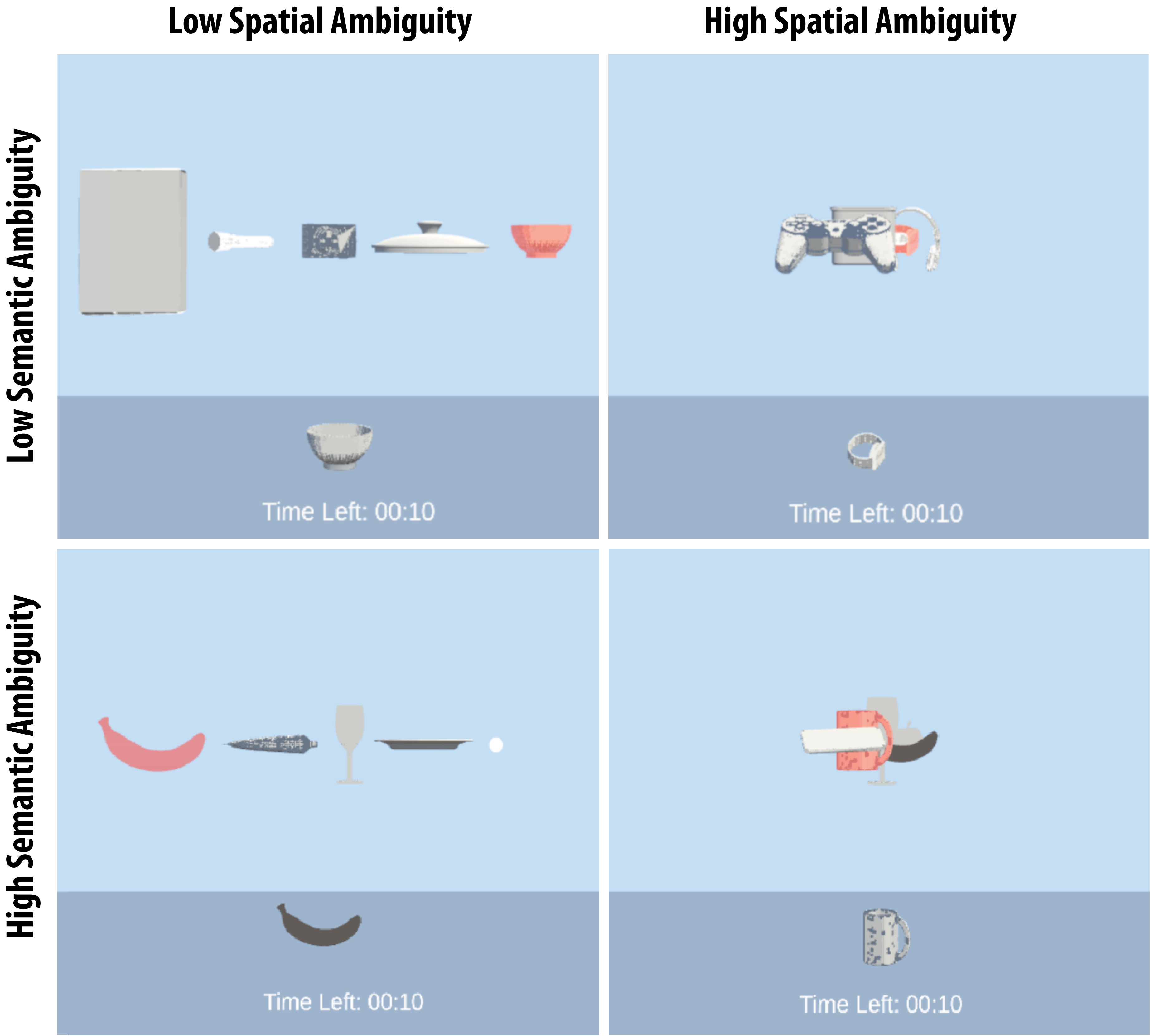}
    \caption{Example scenes illustrating the four ambiguity conditions that vary along semantic and spatial dimensions. In the low semantic × low spatial condition, objects (cracker box, flashlight, camera, disk lid, and bowl) are both semantically and spatially distinct, minimizing potential confusion. The low semantic × high spatial condition (game controller, meat can, flashlight, watch, and headphone) maintains semantic distinctiveness but increases spatial proximity, raising the likelihood of spatial ambiguity. Conversely, the high semantic × low spatial condition (banana, toothpaste, wineglass, plate, and small ball) introduces semantically similar items while keeping them spatially separated. Finally, the high semantic × high spatial condition (smartphone, mug, banana, apple, and wineglass) combines both semantic similarity and spatial proximity, creating the highest level of ambiguity. }
    \Description{The figure shows four example VR scenes arranged in a 2×2 grid, illustrating combinations of semantic and spatial ambiguity.
    Top left (low semantic × low spatial ambiguity): Five objects — cracker box, flashlight, camera, disk lid, and bowl — are distinct and well spaced apart, minimizing confusion.
    Top right (low semantic × high spatial ambiguity): Five semantically distinct objects — game controller, meat can, flashlight, watch, and headphones — are placed close together, raising spatial ambiguity.
    Bottom left (high semantic × low spatial ambiguity): Four semantically similar objects — banana, toothpaste, wineglass, plate, and a small ball — are presented, but spaced apart to reduce spatial ambiguity.
    Bottom right (high semantic × high spatial ambiguity): Five objects — smartphone, mug, banana, apple, and wineglass — combine semantic similarity with close spatial arrangement, producing the highest level of ambiguity.
    }
    \label{fig:scene_layout}
\end{figure}

\subsubsection{Study Design} \label{sec: study_design}
We used a within-subjects design with a $3 \times 2 \times 2$ factorial structure. Independent variables were:
 
\begin{itemize}
    \item \textbf{Interaction Method:} 
   \textsc{Point} (excluding gestural cues), \textsc{Grasp} (excluding directional cues), and \textsc{Point\&Grasp} (our proposed Bayesian integration of both cues).
    \item \textbf{Spatial Ambiguity:} 
    low (1° angular separation between adjacent objects, causing strong occlusion) vs.\ high (5° separation, ensuring no overlap).
    \item \textbf{Semantic Ambiguity:} 
    low (objects from distinct clusters) vs.\ high (three objects from the same cluster and two from different clusters).
\end{itemize}
Crossing semantic and spatial ambiguity yielded four conditions: Low semantic × Low spatial, Low semantic × High spatial, High semantic × Low spatial, and High semantic × High spatial.
Example scenes are shown in Figure~\ref{fig:scene_layout}.
For semantic ambiguity, we prepared 16 object sets in total: 8 low-semantic sets and 8 high-semantic sets, each containing five objects (details in supplementary). For each set, two spatial layouts were instantiated (low vs.\ high). 
Each participant was assigned four low-semantic and four high-semantic sets, with each set appearing in both spatial layouts, resulting in 16 unique scenes per participant.
Within each scene, interaction methods were tested, and the order of methods and semantic ambiguity conditions was counterbalanced across participants.

\subsubsection{Task}
\begin{figure}
    \centering
    \includegraphics[width=0.75\linewidth]{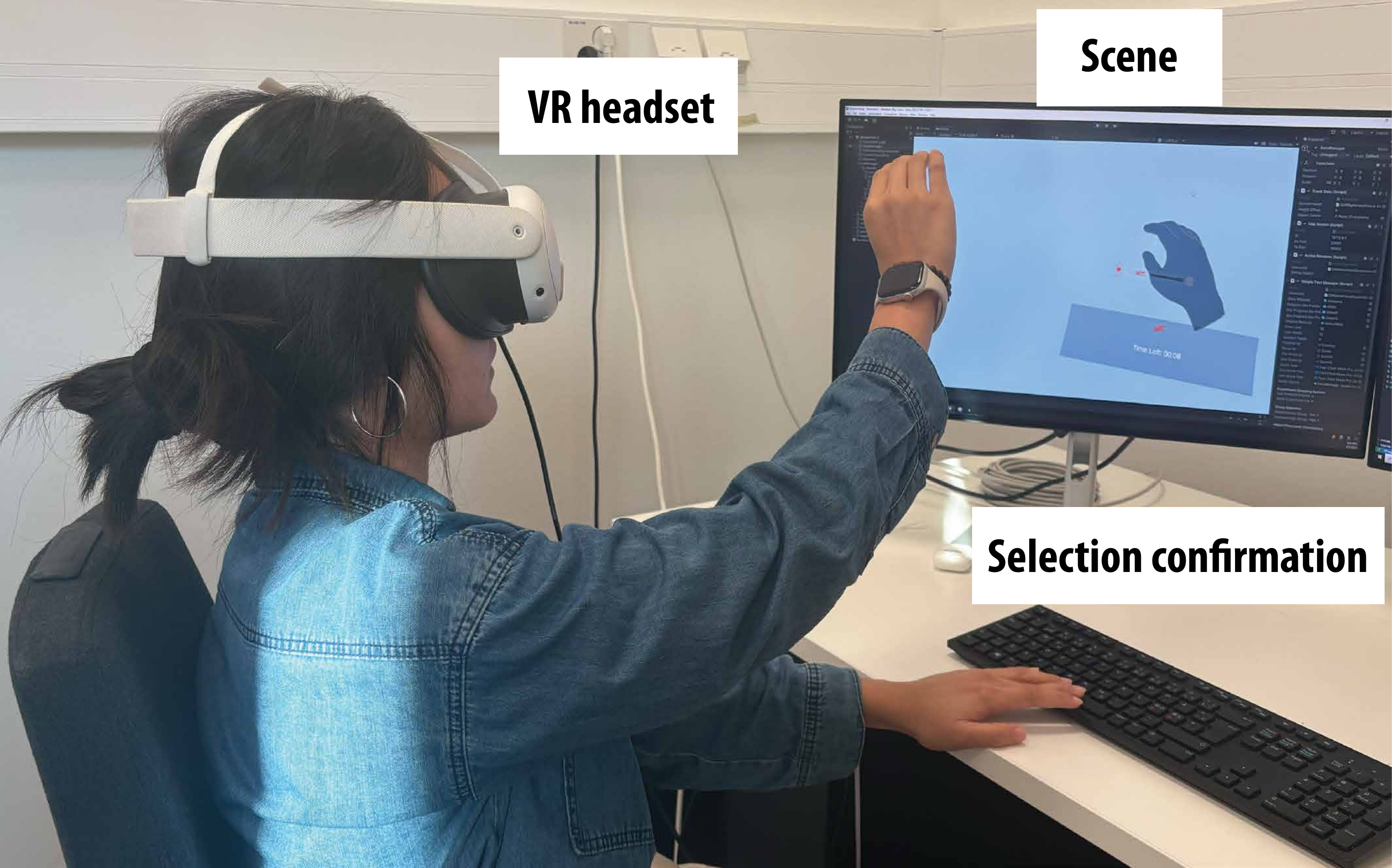}
    \caption{Study setup. Participants wore a VR headset with hand tracking. The scene displayed five candidate objects, with one highlighted in red as the target. Participants provided directional or grasping cues with their right hand and used the space bar with the left hand to confirm the selection.}
    \Description{The figure shows the study setup. A participant is seated at a desk, wearing a VR headset equipped with hand tracking. On the computer monitor in front of them, the VR scene displays a virtual hand and five candidate objects, with one object highlighted in red as the target. The participant uses their right hand to provide cues (directional or grasping gestures) while their left hand rests on the keyboard, pressing the space bar to confirm selections. Labels indicate the VR headset, the scene on the monitor, and the selection confirmation step.}
    \label{fig:study_setup}
\end{figure}

Each trial presented five candidate objects in the VR scene, with one highlighted in red as the target (as illustrated in Figure~\ref{fig:study_setup}). Participants selected the target using the assigned Interaction method and confirmed their choice by pressing the space bar with their left hand.  
Each scene comprised five trials, with each object serving once as the target. Objects were placed 2–3 meters away, and each trial had a 10-second time limit. After incorrect selections, participants could retry until selecting the correct target or reaching the time limit, at which point an auditory cue signaled the end of the trial and prompted them to proceed to the next trial.

\begin{table*}[t]
\centering
\caption{(Study 1) Statistical analysis of selection time and trial completion rate.}
\label{tab:anova}
\resizebox{0.7\textwidth}{!}{%
\begin{tabular}{lccccc}
\toprule
 &  & \multicolumn{2}{c}{Selection time} & \multicolumn{2}{c}{Completion rate} \\
\cmidrule(lr){3-4} \cmidrule(lr){5-6}
 & df & F-ratio & p-value & F-ratio & p-value \\
\midrule
Method & (2, 22) & 13.439 & \bf{<0.001} & 4.754 & \bf{0.035} \\
Spatial Ambiguity & (1, 11) & 165.407 & \bf{<0.001} & 49.677 & \bf{<0.001} \\
Semantic Ambiguity & (1, 11) & 58.537  & \bf{<0.001} & 49.751  & \bf{<0.001} \\
Method $\times$ Spatial Ambiguity & (2, 22) & 83.508  & \bf{<0.001} & 38.436 & \bf{<0.001} \\
Method $\times$ Semantic Ambiguity & (2, 22) & 19.282 & \bf{<0.001} & 19.705  & \bf{<0.001} \\
Spatial Ambiguity $\times$ Semantic Ambiguity & (1, 11) & 5.189  & \bf{0.044} & 0.823  & 0.384 \\
Method $\times$ Spatial Ambiguity $\times$ Semantic Ambiguity & (2, 22) & 1.747  & 0.205 & 2.314  & 0.131 \\
\bottomrule
\end{tabular}}
\end{table*}

\begin{figure*}
    \centering
    \includegraphics[width=0.96\linewidth]{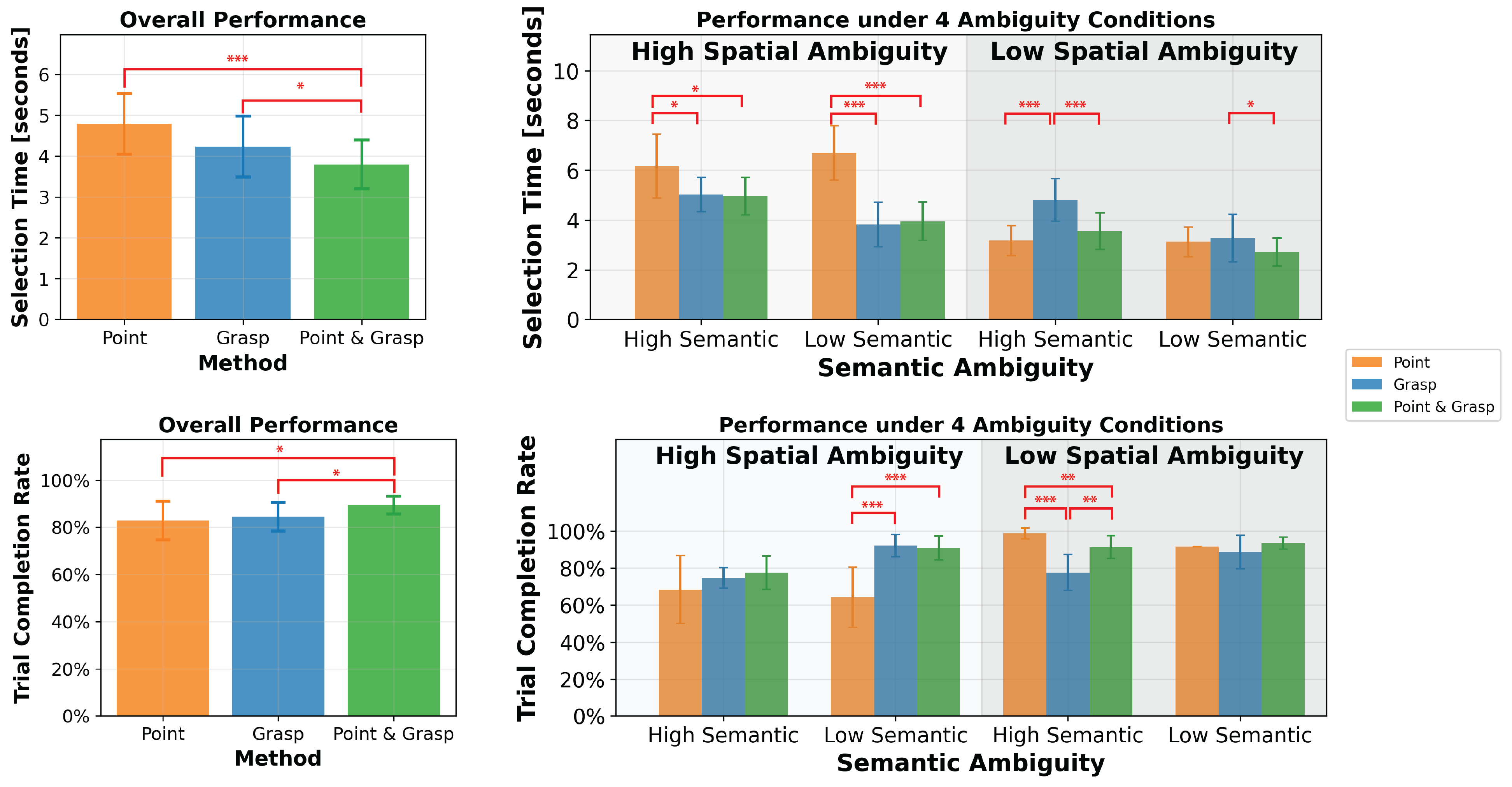}
    \caption{(Study 1) \textsc{Point\&Grasp} achieves the fastest selection times and the highest trial completion rates, avoiding the slowdown of \textsc{Point} under spatial ambiguity and \textsc{Grasp} under semantic ambiguity. Across both measures, \textsc{Point\&Grasp} remains robust under spatial and semantic ambiguity where single-cue methods degrade. Error bars show between-participant variation. Asterisks indicate statistically significant differences after Bonferroni correction (*$p < 0.05$, **$p < 0.01$, ***$p < 0.001$).}

    \Description{The figure shows the selection time and trial completion rate of the three techniques across overall performance and different ambiguity conditions. Overall, Point & Grasp exhibits shorter selection times and higher completion rates than the single-cue methods. When performance is broken down by spatial and semantic ambiguity, Point is mainly affected by increased spatial ambiguity, while Grasp degrades under high semantic ambiguity. In contrast, Point & Grasp shows relatively consistent performance across all conditions, indicating reduced sensitivity to either ambiguity source. Variability across participants is reflected by the error bars, and statistically significant differences between methods are indicated by asterisks.}
    \label{fig:result_study1_performance}
\end{figure*}

\subsubsection{Procedure}
Upon arrival, participants provided informed consent and received a briefing about the study. They then practiced all three interaction methods with sample objects until comfortable.
In the main study, each participant completed three sessions, one per interaction method, with session order counterbalanced. Within each session, they performed 16 scenes in randomized order, yielding five trials per scene. Participants could rest between scenes or sessions as needed.
After finishing all sessions, participants filled out a brief questionnaire assessing the learning difficulty of the \textsc{Grasp} method, the naturalness of the gestures they produced compared to everyday grasping, and the ease of learning \textsc{Point\&Grasp}.

\subsubsection{Measures}
We measured two objective performance metrics: target selection time (from trial onset to correct selection or the 10-second limit) and trial completion rate (the percentage of trials in which the target was correctly selected within the time limit). For each trial, we also logged the system’s cue probabilities to analyze how directional and gestural evidence contributed to target inference, particularly in \textsc{Point\&Grasp}.

In addition, participants completed a post-study questionnaire providing subjective ratings (1) the ease of learning grasp-based selection, (2) the naturalness of the produced grasping gestures compared to everyday grasping, and (3) the ease of learning \textsc{Point\&Grasp}, all on a 5-point Likert scale.

\begin{figure*}
    \centering
    \includegraphics[width=1.0\linewidth]{figures/combined_score_selection_patterns.png}
    \caption{(Study 1) Users flexibly exploited gesture and direction cues in \textsc{Point\&Grasp}, with successes often driven by a single dominant cue but still possible when neither cue was sufficient alone. Trials were labeled Both (both cues selected the target), Gesture Only (only gesture did), Direction Only (only direction did), or Neither (neither did). Left: overall patterns (success vs. failure). Right: breakdown by ambiguity conditions. Values denote percentages within each bar.}

    \Description{The figure shows stacked bar charts illustrating how users combined cues in the \textsc{Point\&Grasp} condition. On the left, overall results compare successful and failed trials. Most successes involve both cues (gesture and direction) or a single dominant cue, with only a small fraction succeeding when neither cue was sufficient. On the right, results are broken down by four ambiguity conditions. In high spatial plus high semantic ambiguity, direction-only cues dominate failures. In high spatial plus low semantic ambiguity, both cues are often used successfully. In low spatial plus high semantic ambiguity, gesture-only cues account for a substantial portion of successes. In low spatial plus low semantic ambiguity, both cues again dominate successes. Percentages within each bar indicate the proportion of trials attributed to each cue type.}
    \label{fig:combi_agreement_analysis}
\end{figure*}

\subsection{Results}
We organize the results according to these questions. 
Section~\ref{sec: comparison across methods} compares the three Interaction methods, 
Section~\ref{sec: cue agreement} analyzes cue agreement in the \textsc{Point\&Grasp} method, 
and Section~~\ref{sec: subjective evaluation} reports participants’ questionnaire responses.

\subsubsection{Comparison of Methods} \label{sec: comparison across methods}
A three-way (Method × Spatial Ambiguity × Semantic Ambiguity) repeated-measures ANOVA with Greenhouse–Geisser correction revealed significant main effects of Method, Spatial Ambiguity, and Semantic Ambiguity on both selection time and completion rate (Table~\ref{tab:anova}). We also found significant two-way interactions of Method × Spatial Ambiguity and Method × Semantic Ambiguity on both measures, and a smaller but reliable Spatial × Semantic Ambiguity interaction on selection time only. No significant three-way interaction was observed.

\subsubsection{Selection Time.} Figure~\ref{fig:result_study1_performance} (top left) shows the overall selection time across methods. \textsc{Point\&Grasp} was the fastest on average, while \textsc{Point} required the longest completion time. 
Post-hoc tests with Bonferroni correction confirmed that \textsc{Point\&Grasp} was significantly faster than both \textsc{Point} ($p<0.001$) and \textsc{Grasp} ($p=0.016$).
We break down the results by ambiguity condition, given that the ANOVA revealed significant interactions of Method × Spatial Ambiguity and Method × Semantic Ambiguity.
The results confirmed that each single-cue method was affected by a different source of ambiguity.
Under high spatial ambiguity, \textsc{Point} showed markedly longer mean times than both \textsc{Grasp} ($p=0.002$) and \textsc{Point\&Grasp} ($p<0.001$). In contrast, \textsc{Grasp} slowed under high semantic ambiguity, with longer durations than \textsc{Point\&Grasp} ($p=0.007$). 
\textsc{Point\&Grasp}, with the use of the complementary cues, maintains robustly low selection times across conditions.
Pairwise comparisons between methods under each ambiguity condition are presented in the top right panel of Figure~\ref{fig:result_study1_performance}.

\subsubsection{Trial Completion Rate.}
Figure~\ref{fig:result_study1_performance} (bottom left) summarizes the overall completion rates.
\textsc{Point\&Grasp} achieved the highest success rate, significantly outperforming both \textsc{Point} ($p=0.024$) and \textsc{Grasp} ($p=0.030$).
When analyzed by ambiguity condition (Figure~\ref{fig:result_study1_performance}, bottom right), the results once again highlight the complementary strengths of the different modality cues.
Under high spatial ambiguity, \textsc{Point} performance degraded substantially, resulting in significantly lower completion rates than both \textsc{Grasp} ($p=0.002$) and \textsc{Point\&Grasp} ($p<0.001$).
In contrast, under high semantic ambiguity, \textsc{Grasp} performance dropped, yielding significantly lower completion rates than \textsc{Point\&Grasp} ($p=0.007$), while \textsc{Point} remained relatively more robust.
Across all conditions, \textsc{Point\&Grasp} again consistently delivered the highest completion rates.
These findings mirror the results with selection time, reinforcing that cue integration provides robustness against both spatial and semantic sources of uncertainty.

\subsubsection{Cue Agreement Analysis of \textsc{Point \& Grasp} Method} \label{sec: cue agreement}
To unpack the mechanisms behind \textsc{Point\&Grasp}’s performance, we examined how agreement between the two cues related to task outcomes. For each trial, we checked whether the gestural and directional cues independently ranked the target the highest and grouped trials into four categories: Both, Gesture Only, Direction Only, and Neither. This analysis reveals whether robustness stems mainly from cue convergence or from the system’s ability to rely on whichever cue is informative in context.

As illustrated in Figure~\ref{fig:combi_agreement_analysis}, successes were dominated by cases where a single cue alone identified the target—Gesture Only accounted for 48.9\% of successes, compared with 15.3\% Direction Only and 26.3\% Both. Neither was rare (9.6\%), yet even in these cases the fused model sometimes recovered the correct target, demonstrating that  \textsc{Point\&Grasp}’s robustness does not rely on cue agreement; rather, it hinges on the ability to capitalize on whichever cue provides strong evidence in context, and at times to accumulate weaker signals from both cues. Failures, by contrast, were largely driven by Neither (68.9\%), indicating that errors most often occurred when both cues provided weak evidence.

\begin{figure*}[h]
    \centering
    \includegraphics[width=\linewidth]{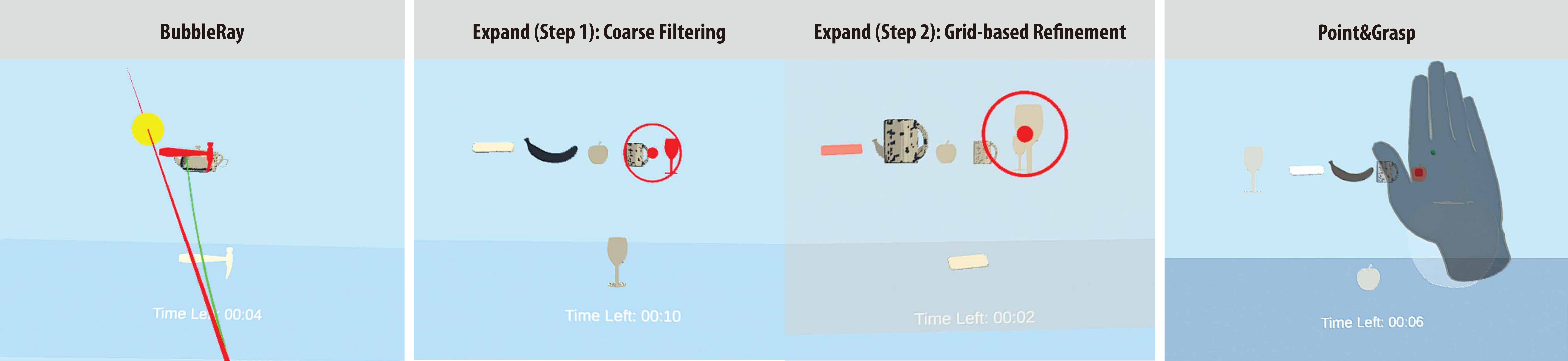}
    \caption{Evaluated techniques in Study 2. 
    \textbf{BubbleRay:} A red ray is cast from the hand; the yellow disc marks the bubble tangent to the selected object; the green curve indicates the chosen target.
    \textbf{Expand (Step 1 \& Step 2):} Coarse selection is performed by using a circular cursor to gather nearby objects; the collected candidates are then arranged into a screen-space grid for refined disambiguation.
    \textbf{\textsc{Point\&Grasp}:} The green endpoint cursor provides directional feedback. The red cursor and semi-transparent target visualization indicate the selected object. A flat-hand posture, regardless of palm orientation, is used here as a ``null'' gesture, which the model interprets as yielding a uniform gesture--object likelihood.}
    \Description{This figure compares three 3D selection techniques evaluated in Study 2: BubbleRay, Expand, and Point&Grasp. The figure contains four panels arranged from left to right. 1. BubbleRay: A red ray extends from the user’s hand toward the scene. A yellow circular disc marks the tangent point where the ray touches an object. A green curve indicates the object selected by the system. Several household objects (such as a wine glass, a banana, and small items) appear in the scene. 2 . Expand – Step 1 (Coarse Filtering): A circular selection cursor surrounds nearby objects to gather them as candidates. Several objects lie within the circular boundary, which is shown in red. Objects outside this boundary are excluded at this stage. Expand – Step 2 (Grid-based Refinement): The coarse candidates identified in Step 1 are rearranged into a 2D screen-space grid. Each grid cell contains one candidate object. A red highlight around one grid cell shows the user’s refined selection. 3. Point&Grasp: A virtual hand model is shown from a first-person perspective. A green endpoint cursor at the hand provides directional feedback. A red dot marks the object the system currently interprets as the gesture-based target. A semi-transparent sphere indicates the selected object. The example also includes a “null gesture,” in which a flat-hand posture yields a uniform gesture–object likelihood.}
    \label{fig:sota_intro}
\end{figure*}

Breaking down by ambiguity reveals when each cue is most informative. Under high spatial $\times$ low semantic ambiguity, successes were overwhelmingly Gesture Only (67.4\%), with Direction Only contributing little (6.9\%): spatial crowding undermines pointing, making grasp-compatibility the decisive signal. Under low spatial $\times$ high semantic ambiguity, contributions became balanced—Gesture Only 30.1\%, Direction Only 27.9\%, and Both 28.8\%—highlighting that pointing regains discriminative power when objects are well separated but semantically similar. In the most difficult high spatial $\times$ high semantic condition, successes still leaned heavily on Gesture Only (66.1\%), whereas failures were mostly Neither (74.1\%), reflecting simultaneous cue degradation. Finally, under low–low ambiguity, Both peaked (35.9\%), indicating frequent cue convergence and near-ceiling performance.

Overall, these patterns clarify \textsc{Point\&Grasp}’s advantage: Baye\-sian integration adapts by leaning on the cue that is informative---gesture under spatial ambiguity, direction under semantic ambiguity---and amplifies confidence when cues agree. This adaptive behavior directly explains the consistently faster selection times and higher completion rates observed in Section \ref{sec: comparison across methods}.


\subsubsection{Subjective Evaluation} \label{sec: subjective evaluation}
Post-study questionnaire results indicate that participants found grasping gestures easy to learn ($M = 1.92$, $SD = 0.51$, 5-point scale, 1 = very easy, 5 = very hard) and fairly natural compared to everyday grasping habits ($M = 3.67$, $SD = 0.65$, 5-point scale, 1 = not at all, 5 = very matching). Participants also reported that \textsc{Point\&Grasp} was very easy to learn ($M = 1.08$, $SD = 0.28$). 
Overall, these subjective findings align with the objective results, suggesting that grasping gestures are both intuitive and learnable, and can be smoothly integrated with pointing cues to support effective out-of-reach object selection.

\section{Study 2: Comparing with State-of-the-Art 3D Selection Techniques} \label{sec: study_2}
While \hyperref[sec: study_1]{Study~1} examined the benefits of cue integration by comparing \textsc{Point\&Grasp} with its single-cue variants, \hyperref[sec: study_2]{Study~2} evaluates its performance against state-of-the-art 3D selection techniques. 
We selected two empirically strong methods from prior work—BubbleRay~\cite{lu2020investigating} and Expand~\cite{cashion2012dense}—which both rely on directional cues but employ distinct mechanisms to mitigate their limitations in dense target scenarios.

\subsection{Method}

\subsubsection{Interaction Methods} \label{sec: study2_interaction_methods}

We evaluated three interaction methods, illustrated in Figure~\ref{fig:sota_intro}.

\begin{itemize}
    \item \textbf{BubbleRay}~\cite{lu2020investigating}. It performs immediate selection by dynamically shaping a bubble region around the pointing ray (inspired by the original Bubble Cursor~\cite{grossman2005bubble}), and selects the object closest to the bubble boundary. 
    Among the variants described by the authors, we implemented the version based on angular distance between the ray and candidate objects. 
    This variant has been shown to outperform a wide range of established directional techniques, including 3D BubbleCursor~\cite{vanacken2007exploring}, Go-Go~\cite{poupyrev1996go}, and SQUAD-style ray-based methods~\cite{kopper2011rapid}, making it a strong representative of state-of-the-art directional selection.
    
    \item \textbf{Expand}~\cite{cashion2012dense}. It adopts a two-stage selection strategy consisting of coarse filtering followed by grid-based refinement. During coarse selection, a circular cursor gathers nearby objects (in our setting, within a 10 cm Euclidean distance threshold) around the ray. The collected candidates are then arranged into a 2D screen-space grid to support disambiguation. 
     Prior work has shown Expand to be significantly faster and more accurate than SQUAD~\cite{kopper2011rapid} and Zoom~\cite{cashion2012dense}. More recently, in densely arranged 3D selection scenarios, a grid-based method inspired by Expand has also outperformed other contemporary baselines~\cite{yu2020fully}.
    
    \item \textbf{\textsc{Point\&Grasp} (Ours)}. For Study~2, we introduced two refinements: (1) We added a circular cursor at the ray endpoint to ensure consistent directional feedback as with BubbleRay and Expand. 
    Prior work suggests that explicit cursor representations help anchor user attention and reduce perceptual ambiguity~\cite{lin2026interaction}. 
    (2) We retrained the gesture model to include a ``null'' flat-hand gesture that conveys no grasp semantics. 
    This gesture produces a uniform likelihood on objects and 
    and enables the system to correctly interpret cases where users do \emph{not} intend to provide a semantic cue, allowing directional evidence to dominate posterior inference.
\end{itemize}

To ensure a fair comparison, all methods used the same definition of ray direction as described in Section~\ref{sec: directional_cue}.

\begin{table*}[]
\centering
\caption{(Study 2) Statistical analysis of selection time and trial completion rate.}
\label{tab:anova_study2}
\resizebox{0.7\textwidth}{!}{%
\begin{tabular}{lccccc}
\toprule
 &  & \multicolumn{2}{c}{Selection time} & \multicolumn{2}{c}{Completion rate} \\
\cmidrule(lr){3-4} \cmidrule(lr){5-6}
 & df & F-ratio & p-value & F-ratio & p-value \\
\midrule
Method & (2, 22) & .070 & .885 & 27.132 & \bf{<0.001} \\
Spatial Ambiguity & (1, 11) & 248.553 & \bf{<0.001} & 138.078 & \bf{<0.001} \\
Semantic Ambiguity & (1, 11) & 6.227  & \bf{0.030} & 1.510  & 0.245 \\
Method $\times$ Spatial Ambiguity & (2, 22) & 44.936  & \bf{<0.001} & 32.783 & \bf{<0.001} \\
Method $\times$ Semantic Ambiguity & (2, 22) & 8.384 & \bf{0.008} & 14.018  & \bf{<0.001} \\
Spatial Ambiguity $\times$ Semantic Ambiguity & (1, 11) & 16.486  & \bf{<0.001} & 22.172  & \bf{<0.001} \\
Method $\times$ Spatial Ambiguity $\times$ Semantic Ambiguity & (2, 22) & 16.486  & \bf{<0.001} & 22.172 & \bf{<0.001} \\
\bottomrule
\end{tabular}}
\end{table*}

\begin{figure*}
    \centering
    \includegraphics[width=1.0\linewidth]{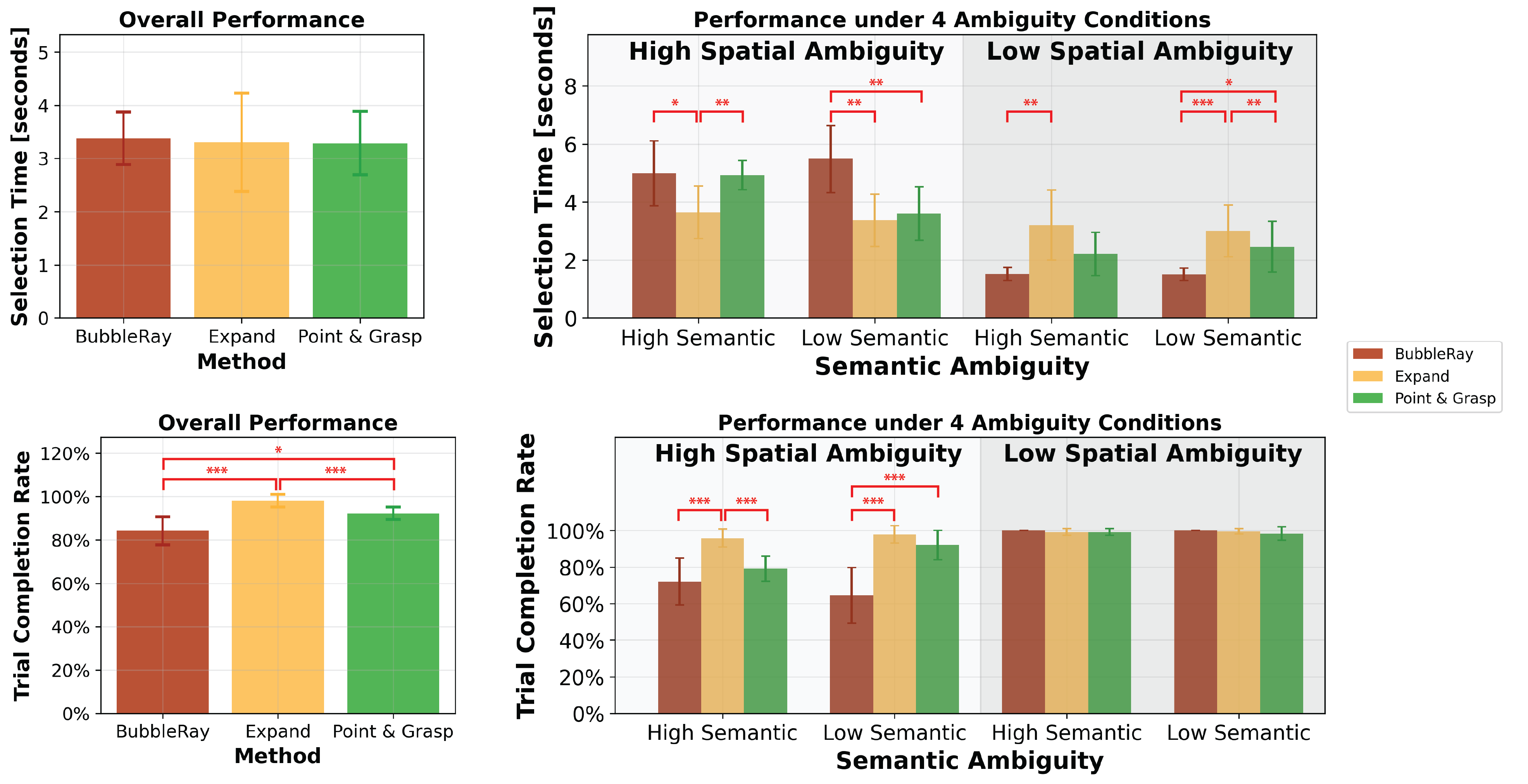}
    \caption{(Study 2) \textsc{Point\&Grasp} achieves faster selection times than Expand under low spatial ambiguity and than BubbleRay under high spatial ambiguity. It also reaches near-Expand completion rates while significantly exceeding BubbleRay, particularly under low semantic ambiguity. Across both measures, \textsc{Point\&Grasp} demonstrates competitive performance against state-of-the-art techniques under varying ambiguity conditions. Error bars show between-participant variation. Asterisks indicate statistically significant differences after Bonferroni correction (*$p < 0.05$, **$p < 0.01$, ***$p < 0.001$).}

    \Description{It presents selection time and trial completion rate for BubbleRay, Expand, and \textsc{Point\&Grasp}, both overall and across combinations of spatial and semantic ambiguity. Overall performance shows comparable selection times across the three techniques, while Expand and \textsc{Point\&Grasp} achieve higher completion rates than BubbleRay. When broken down by ambiguity conditions, BubbleRay and Expand exhibit longer selection times under high spatial ambiguity, whereas \textsc{Point\&Grasp} remains relatively stable. Completion rates reveal a similar pattern: BubbleRay degrades notably under high spatial ambiguity, while Expand and \textsc{Point\&Grasp} maintain high success, particularly when semantic ambiguity is low. Error bars indicate between-participant variation, and asterisks denote statistically significant differences after Bonferroni correction.}
    \label{fig:result_study2_sota}
\end{figure*}

\subsubsection{Apparatus and Participants}
The apparatus was identical to that used in Study~1, including the Meta Quest~2 headset connected via Oculus Link and the same desktop computer. 
We recruited a new set of twelve participants (5 male, 7 female; aged 25--35) from the university community. 
All participants had normal or corrected-to-normal vision and were right-handed.

\subsubsection{Study Design}
Study~2 adopted the same within-subjects $3 \times 2 \times 2$ factorial design as \hyperref[sec: study_1]{Study~1}, with identical independent variables: Interaction Method, Spatial Ambiguity, and Semantic Ambiguity. The only difference lies in the levels of the Interaction Method factor, which in Study~2 compared \textsc{Point\&Grasp} against two state-of-the-art directional techniques, BubbleRay and Expand. 
Spatial and Semantic Ambiguity were carried over unchanged, resulting in the same four ambiguity configurations shown in Figure~\ref{fig:scene_layout}. 
Details on scene construction, object sets, and ambiguity manipulation are provided in Section~\ref{sec: study_design}.

\subsubsection{Task, Procedure, and Measures}

The task and procedure were identical to those in Study~1, except that no subjective ratings were collected in Study~2. 
We evaluated the three Interaction methods with \textit{target selection time} and \textit{trial completion rate}. 
For full details of the task and procedure, please refer to Section~\ref{sec: study_1}.

\subsection{Results}


A three-way repeated-measures ANOVA (Method $\times$ Spatial Ambiguity $\times$ Semantic Ambiguity) revealed a significant main effect of Spatial Ambiguity on both selection time and completion rate, a significant main effect of Method on completion rate but not selection time, and a significant main effect of Semantic Ambiguity on selection time only (Table~\ref{tab:anova_study2}). 
Significant two-way interactions were observed for Method $\times$ Spatial Ambiguity, Method $\times$ Semantic Ambiguity, and Spatial $\times$ Semantic Ambiguity, as well as a significant three-way interaction for both outcome measures. Bonferroni-corrected post hoc tests were conducted for all significant effects, and the results reported below reflect these follow-up analyses.

\subsubsection{Selection Time.}
Figure~\ref{fig:result_study2_sota} (top left) shows no significant Method effect, with all techniques exhibiting similar overall speed.
Condition-specific comparisons, however, revealed clear differences.
Between the two state-of-the-art baselines, Expand was faster under high spatial ambiguity ($p=0.003$), whereas BubbleRay was faster under low spatial ambiguity ($p=0.001$), reflecting the trade-off between refinement overhead and robustness in occluded layouts.
\textsc{Point\&Grasp} consistently fell between the two: it was faster than Expand under low spatial ambiguity ($p=0.010$) and faster than BubbleRay under high spatial ambiguity ($p=0.035$).
Importantly, semantic ambiguity further highlighted \textsc{Point\&Grasp}’s advantage. 
Under low semantic ambiguity---when gesture cues become reliable and informative---it showed significantly faster performance ($p=0.013$) and maintained stable speed across spatial conditions.
For example, under the low-semantic $\times$ high-spatial condition, \textsc{Point\&Grasp} achieved selection speed comparable to Expand, but surpassed it under the low-semantic $\times$ low-spatial condition ($p=0.004$).


\subsubsection{Trial Completion Rate.}
Figure~\ref{fig:result_study2_sota} (bottom left) shows the overall completion rates across the three techniques. 
Expand achieved the highest accuracy, followed by \textsc{Point\&Grasp}, with BubbleRay performing least accurately. 
Breaking the results down by ambiguity condition clarifies the source of Expand’s advantage. 
Under low spatial ambiguity, all three techniques performed near ceiling (i.e., 100\%), showing little separation. 
Under high spatial ambiguity, however, Expand’s refinement step provided a clear benefit, yielding substantially higher completion rates than the other two methods ($p < 0.001$ for both).
For \textsc{Point\&Grasp}, performance again varied with semantic ambiguity ($p=0.006$): under the high semantic $\times$ high spatial condition,\textsc{Point\&Grasp} performed significantly worse than Expand ($p < 0.001$), whereas under low semantic ambiguity it achieved completion rates comparable to Expand (no significant difference) and significantly outperformed BubbleRay ($p < 0.001$).

\subsubsection{Summary.}
The results of Study~2 suggest that \textbf{multi-cue integration reduces \textsc{Point\&Grasp}’s dependence on the spatial-ambiguity dimension by incorporating semantic information.} 
BubbleRay and Expand showed a clear trade-off along this spatial dimension---BubbleRay performing better in low-spatial scenes and Expand excelling in high-spatial scenes.
However, when semantic cues were reliable (i.e., under low semantic ambiguity), \textsc{Point\&Grasp} demonstrated strong robustness to spatial interference.
It matched Expand’s speed and accuracy under high spatial ambiguity, while avoiding Expand’s refinement overhead in low spatial ambiguity, resulting in faster performance in those layouts. 
This highlights how incorporating a complementary cue dimension enables \textsc{Point\&Grasp} to maintain stable and reliable performance where state-of-the-art directional-cue methods exhibit trade-offs.

\section{Discussion}

Both studies demonstrate clear benefits of probabilistic cue integration for out-of-reach object selection.
Based on these findings, the key features of \textsc{Point\&Grasp} can be summarized as follows:

\begin{itemize}
    \item \textbf{Overall performance:}
     In \hyperref[sec: study_1]{Study~1}, \textsc{Point\&Grasp} surpassed both single-cue baselines in speed and accuracy. In \hyperref[sec: study_2]{Study~2}, it achieved performance comparable to state-of-the-art directional techniques, consistently landing between BubbleRay and Expand, which are in trade-off along spatial ambiguity.
     
    \item \textbf{Robustness under different ambiguity:}  
    The advantage of cue integration became evident under different sources of ambiguity.
    \hyperref[sec: study_1]{Study~1} showed that \textsc{Point} and \textsc{Grasp} degrade under different ambiguity sources, while \textsc{Point\&\-Grasp} remained stable across both. 
    \hyperref[sec: study_2]{Study~2} further demonstrated that, compared to state-of-the-art techniques, multi-cue integration reduces \textsc{Point\&\-Grasp}’s dependence on the spatial-ambiguity dimension by incorporating semantic information.
    Under low semantic ambiguity, \textsc{Point\&Grasp} maintained robust performance even in highly cluttered scenes, matching Expand in high-spatial layouts, while avoiding its refinement overhead in low-spatial layouts.
    
    \item \textbf{Context-dependent cue dominance:}  As directional and grasping gestural cues are complementary, the relative contribution of each changes with the type of ambiguity. Bayesian cue integration naturally lets the more discriminative cue dominate. The cue agreement analysis showed that the dominance of grasping cues increased as scenes became more spatially ambiguous and less semantically ambiguous.
    
    \item \textbf{Probabilistic rescue:} The Bayesian framework enabled successful selection even when neither cue alone maximized likelihood. Approximately 10\% of \hyperref[sec: study_1]{Study~1} trials fell into this category, showing that probabilistic integration extends performance beyond the limits of single-cue methods.
    
    \item \textbf{Real-time operation:} All probabilistic computations, including the deployed gesture-object likelihood model, were lightweight enough to support real-time object selection, ensuring practical applicability in MR environments.
\end{itemize}
Below, we discuss broader implications and open questions.
    

\subsubsection*{Are in-reach grasping datasets applicable to out-of-reach interaction?}
The \dataset~dataset reveals statistically significant differences between in-reach and out-of-reach grasping gestures performed by the same users on the same objects (see Section~\ref{sec:gesture_analsysis}).
In out-of-reach conditions, fingertip positions shifted by around 3~cm and grasp apertures became noticeably narrower---likely reflecting perceptual and motor uncertainties when imagining grasps on distant objects.
These deviations indicate that directly applying physical in-reach grasp datasets such as GRAB to out-of-reach interaction can introduce systematic bias in gesture–object likelihood modeling.

\subsubsection*{How does multi-cue integration shape user strategies under ambiguity?}
Our cue agreement analysis in \hyperref[sec: study_1]{Study~1} (see Section.~\ref{sec: cue agreement}) suggests that users may adaptively shift their selection strategy depending on scene ambiguity. 
Under high spatial ambiguity, around 67\% of successful trials were driven by gesture cues, but this proportion dropped to 30\% in low-ambiguity settings. 
These shifts reflect not only changes in cue informativeness with scene context but also users strategically privileging the cue that best reduces uncertainty. 
When spatial crowding made directional cues unreliable, they leaned on grasp cues; when objects were semantically similar but spatially distinct, users adopted a more balanced strategy that leveraged both directional and grasp information.

\subsubsection*{How can users recover from mis-selections?}
Although probabilistic cue integration allows \textsc{Point\&Grasp} to handle cases where neither cue alone is decisive, \hyperref[sec: study_2]{Study~2} reveals substantial room for user-driven recovery.
The two refinements introduced in \hyperref[sec: study_2]{Study~2}—an endpoint cursor for clearer directional feedback and a null gesture to intentionally disable the gestural cue—significantly reduced mis-selections, raising completion rates in low-ambiguity scenes to near 100\%.
More broadly, because the Bayesian model maintains posterior probabilities over all candidates, our probabilistic cue integration offers an opportunity for additional corrective mechanisms: the system can detect when posterior scores are ambiguous between candidates and trigger lightweight refinement steps only when needed (e.g., a brief confirmatory step within the candidates).
Such mechanisms offer a practical way to resolve semantic and spatial ambiguities among nearby objects in real-world use.

\subsubsection*{What additional modalities could be considered?}
Beyond pointing direction and grasping gestures, our probabilistic framework can be extended with other modalities that complement user intent.
Gaze pointing is a natural candidate to combine with hand pointing, often used to mitigate eye–hand misalignment in VR; similar to directional cues from the hand, gaze endpoints can be modeled as Gaussian likelihoods over candidate objects~\cite{wei2023predicting}.
Speech provides an orthogonal cue, where verbal references to objects can be modeled as categorical likelihoods over candidates whose labels or attributes match the utterance. 
This idea dates back to Bolt’s “Put-that-there” system~\cite{bolt1980put}, which demonstrated the power of combining speech with deictic gestures for object specification.
Finally, action history may offer contextual priors: while not specific to MR, routine modeling work~\cite{banovic2016modeling} suggests that past behavior can inform predictions of future actions, and in our setting, such history could slightly bias the prior toward previously selected objects.

\subsubsection*{How can \textsc{Point\&Grasp} transition from controlled experiments to real MR use?}
Our study employed controlled design choices to investigate the benefits of cue integration in isolation, but each can extend naturally to real MR scenarios.

First, we provide which item to grasp via object-indicating visualization (see Figure~\ref{fig:scene_layout}). Real MR environments include comparable situations---both when users can locate a target through visual search and when they cannot see all details yet still know the intended object (e.g., reaching for a screwdriver to tighten a bolt in a cluttered toolbox). In such cases, users can recall the object’s shape and produce a meaningful grasp without explicit visualization. 
Second, for directional cue, we fixed $\sigma_d$ for evaluation. Real systems can scale it by object size, simply adjusting directional dispersion without altering the fusion mechanism.
Third, target objects were placed at a fixed depth to focus on semantic grasp forms. In real MR settings, users can perceive distance and infer the geometry of familiar objects; nevertheless, viewing distance may shape grasp articulation---through perceived scale or required hand aperture. Investigating these effects remains valuable future work.
Finally, we used a space-bar press for controlled confirmation, but in real MR settings this can be replaced by native gestures such as a left-hand controller click or a bare-hand pinch gesture, allowing the interaction to remain fully \emph{controller-free}.


\subsubsection*{Future work}
Several directions remain open for future research. 
First, the gesture–object likelihood model is limited by the size and diversity of our dataset. Collecting richer gesture data across more objects and users would improve generalization to unseen objects and grasping styles. 
Second, personalization remains an open challenge, as users differ in how they balance directional and gestural cues. Adaptive cue weighting through brief calibration or online adjustment could better align the system with individual strategies. 
Finally, although this work focuses on selection, the same probabilistic framework could extend to downstream MR tasks such as object manipulation, cooperative handover, or multi-user collaboration, where ambiguity is also pervasive.

\section{Conclusion} 
In this paper, we introduced a probabilistic framework for out-of-reach object selection in MR that integrates directional and grasping gestural cues via Bayesian inference.
To support this approach, we built a dataset of out-of-reach grasping gestures and developed a gesture–object likelihood model.
Our evaluation showed that a single user motion can simultaneously express complementary yet separate cues---pointing direction and grasping gesture---that, when fused probabilistically, resolve both spatial and semantic ambiguity. 
By leveraging such complementary cues, our proposed \textsc{Point\&Grasp} method achieved higher accuracy and speed than single-cue baselines consistently under the different ambiguities.
We further showed that, by incorporating semantic information, \textsc{Point\&Grasp} achieves more robust performance across diverse spatial layouts compared with empirically strong directional techniques.
Our result contributes to a line of work in HCI showing the benefits of addressing multimodal input as a probabilistic inference problem.
While prior work has shown strong results in areas like text entry where there is a single end-effector, in this work we have found out a way to exploit the same principle in out-of-reach selection, where one needs to deal with the full kinematic chain of the hand. 
We envision our insights to pave the way to more expressive selection techniques that better exploit the information humans can express with their body.


 

\begin{acks}
This work was supported by the Research Council of Finland (FCAI, 328400, 345604, 341763; Subjective Functions, 357578), the ERC Advanced Grant (101141916), and the National Research Foundation of Korea (RS-2025-00521470). We thank Yutong Du for assistance with early dataset collection and Jiayi He for support in developing the initial VR object selection prototype.
\end{acks}

\bibliographystyle{ACM-Reference-Format}
\balance
\bibliography{references}



\end{document}